\documentclass[letterpaper,titlepage,11pt]{article}
\usepackage{latexsym,amssymb,amstext,amsmath}
\usepackage{slashed}
\usepackage{mathrsfs}
\usepackage{color}
\usepackage{enumerate}
\usepackage{graphicx}
\usepackage{tikz}
\usepackage{etex}
\usepackage{latexsym}
\usepackage{cite}
\allowdisplaybreaks
\newcommand{\ft}[2]{{\textstyle\frac{#1}{#2}}}
\usepackage[
colorlinks=true,
linkcolor=blue,
urlcolor=red,
filecolor=green,
citecolor=red,
pdfstartview=FitV,
pdftitle={},
pdfauthor={Mehmet Ozkan},
pdfsubject={},
pdfkeywords={},
pdfpagemode=None,
bookmarksopen=true
]{hyperref}

\newcommand{\bea}{\setlength\arraycolsep{2pt} \begin{eqnarray}}
\newcommand{\eea}{\end{eqnarray}}
\newcommand{\nn}{\nonumber}

\def\mass{\footnotesize{\textrm M}}

\usepackage{hyperref}

\def\rmi{{\rm i}}
\newsavebox{\uuunit}
\sbox{\uuunit}
{\setlength{\unitlength}{0.825em}
	\begin{picture}(0.6,0.7)
	\thinlines
	\put(0,0){\line(1,0){0.5}}
	\put(0.15,0){\line(0,1){0.7}}
	\put(0.35,0){\line(0,1){0.8}}
	\multiput(0.3,0.8)(-0.04,-0.02){12}{\rule{0.5pt}{0.5pt}}
	\end {picture}}

\newcommand{\U}{\mathop{\rm U}}

\def\be{\begin{equation}}
\def\ee{\end{equation}}
\def\ba{\begin{array}}
\def\ea{\end{array}}
\def\bea{\begin{eqnarray}}
\def\eea{\end{eqnarray}}
\def\bd{\begin{displaymath}}
\def\ed{\end{displaymath}}

\def\nn{\nonumber}

\def\a{\alpha}
\def\b{\beta}
\def\g{\gamma}
\def\G{\Gamma}
\def\d{\delta}

\def\e{\epsilon}

\def\f{\phi}
\def\vf{\varphi}

\def\p{\psi}
\def\P{\Psi}

\def\l{\lambda}
\def\L{\Lambda}
\def\m{\mu}
\def\n{\nu}
\def\r{\rho}
\def\s{\sigma}

\def\t{\tau}

\def\o{\omega}
\def\O{\Omega}
\def\x{\xi}

\def\nn{\nonumber}
\def\cD{\mathcal{D}}
\def\cN{\mathcal{N}}

\def\cL{\mathcal{L}}

\def\cV{\mathcal{V}}

\makeatletter
\@addtoreset{equation}{section}
\makeatother

\begin{document}
%%%%%%%%%%%%%%%%%%%%%%%%%%%%%%%%%%%%%%%%%%%%%%%%%%%%%%%%%%%%%%%%%%%
%
\begin{titlepage}
%%%%%%%%%%%%%%%%%%%%%%%%%%%%%%%%%%

%%%%%%%%%%%%%%%%%%%%%%%%%%%%%%%%%%
\bigskip

\begin{center}
  {\LARGE\bfseries Scale  Invariance in Newton-Cartan and Ho\v{r}ava-Lifshitz Gravity}
  \\[10mm]
\textbf{Deniz Olgu Devecio\u{g}lu$^{1}$, Nese \"Ozdemir$^2$, Mehmet Ozkan$^2$ and  Utku Zorba$^2$}\\[5mm]
\vskip 25pt

{\em $^1$ \hskip -.1truecm Research Institute for Basic Science, Sogang University, \\
	Seoul, 121-742, Korea  \vskip 5pt }

{email: {\tt dodeve@gmail.com}} \\

\vskip 15pt

{\em $^2$ \hskip -.1truecm Department of Physics, Istanbul Technical University,  \\
		Maslak 34469 Istanbul, Turkey  \vskip 5pt }
	
	{email: {\tt nozdemir@itu.edu.tr, ozkanmehm@itu.edu.tr, zorba@itu.edu.tr}}

%{\em  Department of Physics, Istanbul Technical University,  \\
%	Maslak 34469 Istanbul, Turkey, }\\[3mm]
% {\tt ozkanmehm@itu.edu.tr}
\end{center}

\vspace{3ex}

\begin{center}
{\bfseries Abstract}
\end{center}
\begin{quotation} \noindent

We present a detailed analysis of the construction of $z=2$ and $z\neq2$ scale invariant Ho\v{r}ava-Lifshitz gravity. The construction procedure is based on the realization of Ho\v{r}ava-Lifshitz gravity as the dynamical Newton-Cartan geometry as well as a non-relativistic tensor calculus in the presence of the scale symmetry. An important consequence of this method is that it provides us the necessary mechanism to distinguish the local scale invariance from the local Schr\"odinger invariance. Based on this result we discuss the $z=2$ scale invariant Ho\v{r}ava-Lifshitz gravity and the symmetry enhancement to the full Schr\"odinger group.

\end{quotation}

\vfill

%%%%%%%
\flushleft{\today}
%%%%%%
\end{titlepage}
\setcounter{page}{1}
\tableofcontents

\newpage

%%%%%%%%%%%%%%%%%%%%%%%%%%%%%%%%%%%%%%%%%%%%%%%%%%%%%%%%%%%%%%%%%
\section{Introduction}{\label{Intro}}
\paragraph{}
Non-relativistic gravity theories are of considerable interest for numerous reasons that are rooted in their applications in condensed matter and non-relativistic holography. Particularly two noteworthy models, Newton-Cartan and Ho\v{r}ava-Lifshitz gravity, found themselves a wide range of application in recent years. For instance, Newton-Cartan gravity, which is originally developed as the generally covariant description of the Newtonian gravity \cite{Cartan1,Cartan2}, is known be useful in the effective field theory description of the quantum Hall effect \cite{Son:2013rqa,Gromov:2014vla,Geracie:2014nka,Moroz:2014ska}. Ho\v{r}ava-Lifshitz gravity, on the other hand, was proposed for UV completion of General Relativity \cite{Horava1,Horava2} and is useful to study strongly-coupled systems with non-relativistic scaling in the context of non-relativistic holography, see e.g. \cite{Leiva:2003kd,Balasubramanian:2008dm,Son:2008ye,Herzog:2008wg,Duval:2008jg,Kachru:2008yh,Taylor:2008tg, Wang:2017brl}. 

Although these two major examples of non-relativistic gravity have different motivations and formulations, it was shown that there is a dictionary that relates these two theories to each other \cite{Hartong:2015zia}. This dictionary particularly tells us that the  Ho\v{r}ava-Lifshitz gravity can be realized as the dynamical Newton-Cartan geometry by providing a  map between different versions of Ho\v{r}ava-Lifshitz gravity and different torsional extensions of Newton-Cartan geometry. This map then turns the construction of the Ho\v{r}ava-Lifshitz gravity into the construction of invariant higher derivative quantities of  Newton-Cartan geometry with appropriate choice of torsion. Following the usual dictum that more symmetry implies more restriction on the form of the action, a Schr\"odinger tensor calculus, which is the non-relativistic analogue of the conformal tensor calculus, was developed to ease the construction of invariant quantities of Newton-Cartan geometry \cite{Afshar:2015aku}. As a consequence, the conformal extension of the Ho\v{r}ava-Lifshitz gravity was obtained by dictating Schr\"odinger invariance of Newton-Cartan geometry, and was shown that its gauge fixing indeed recovers Ho\v{r}ava-Lifshitz gravity.

The present paper aims to take a step further in this direction by relaxing the Schr\"odinger symmetry to scale symmetry and by developing a scale invariant tensor calculus. This extension is desired for two reasons. First, the Schr\"odinger extension of the Ho\v{r}ava-Lifshitz gravity can only be achieved when $z$, the dynamical critical exponent that characterizes the anisotropy between time and space, is fixed as $z=2$. Thus, the Schr\"odinger tensor calculus developed in \cite{Afshar:2015aku} cannot be applied when $z\neq2$. Scale invariance, on the other hand, can be imposed for arbitrary values of $z$, which makes a scale invariant tensor calculus to have a wider range of applicability for the construction of Ho\v{r}ava-Lifshitz gravity from Newton-Cartan geometry. Second, for $z=2$ theories, the non-relativistic special conformal symmetry dictates that the temporal component of the gauge field of dilatations, $b_\m$ cannot appear in the action \cite{Afshar:2015aku, Bergshoeff:2014uea}. The scale invariant setting has no such constraint, but if the scale invariant actions are combined together in a way that the temporal part of $b_\m$ drops out in the action, then the scale invariance enhances to  Schr\"odinger invariance. Therefore, scale invariant tensor calculus provides a set of actions with a free parameter such that when the free parameter is chosen properly, there is an enhancement to the full Schr\"odinger symmetry.

To develop the framework necessary for the scale invariant tensor calculus, we start with a review of the basics of Newton-Cartan geometry and its torsional extensions in Section \ref{Sec0}. The Newton-Cartan geometry is encoded in a spatial metric $h^{\m\n}$, a temporal vielbein $\t_\m$, and a $\U(1)$ gauge field $m_\m$. An important notion in this geometric setting is that there is no unique way to define the inverse of spatial metric or the temporal vielbein. The definition of the inverse metrics are based on orthogonality relations for the spatial and temporal metrics, but one is always free to choose a different set of inverse metrics in a way that the orthogonality conditions are satisfied. This freedom is known as the Milne boosts and it plays a crucial role in the torsional extension of the Newton-Cartan geometry. Thus, we pay a special attention to the definition of the inverse metrics, their Milne invariance, and the map between different choices of inverse metrics when the Newton-Cartan geometry is extended with torsion. In Section \ref{Section3}, we introduce a scale and Schr\"odinger symmetry to Newton-Cartan geometry. This is achieved by introducing a particular non-metricity condition to the metric compatibility equation. For arbitrary values of $z$, we cannot go beyond the scale symmetry to the full Schr\"odinger extension since a particular Jacobi identity do not close in the presence of Schr\"odinger symmetry unless $z$ is fixed to $z=2$ \cite{Jensen:2014aia}. For $z=2$ the Jacobi identity is satisfied and the last part of this section is devoted to the Newton-Cartan geometry and its torsional extensions in the presence of Schr\"odinger symmetry. In Section \ref{Sec3}, we make the distinction between the scale and the Schr\"odinger invariance explicit by constructing $z=2$ scale invariant models. This is most easily done by gauging the relevant non-relativistic spacetime symmetries. Therefore we separate this section into two main subsections. In the first subsection we give a brief review of Bargmann algebra, which is the central extension of the Galilean algebra, as well as its $z=2$ scale and Schr\"odinger extensions. Based on the gauging of the Bargmann algebra and its extensions, we then construct the relation between the non-relativistic geometric quantities and the group theoretical elements with the relevant symmetry. The second subsection introduces a $z=2$ scale invariant tensor calculus, and includes the construction of the relevant $z=2$ scale invariant geometric quantities. Finally, we show that although our models exhibit scale invariance in general, a particular choice of free parameters lead us to a Schr\"odinger gravity. In Section \ref{Sec4}, we go beyond the $z=2$ theories. To achieve that we repeat procedure that we built in Section \ref{Sec3} and establish the relation between the non-relativistic geometric quantities and the  group theoretical elements of $z\neq2$ scale extended Bargmann algebra. Next, we develop a scale invariant tensor calculus and construct $z\neq2$ non-relativistic gravity models. Finally, using the dictionary between the Newton-Cartan geometry and the Ho\v{r}ava-Lifshitz gravity \cite{Hartong:2015zia}, we construct the $z\neq2$ scale extension of the Ho\v{r}ava-Lifshitz gravity, which is one of the major results of this work. We conclude in Section \ref{Conc}.

%%%%%%%%%%%%%%%%%%%%%%%%%%%%%%%%%%%%%%%%%%%%%%%%%%%%%%%%%%%%%%%%%
\section{Newton-Cartan Gravity}{\label{Sec0}}
%%%%%%%%%%%%%%%%%%%%%%%%%%%%%%%%%%%%%%%%%%%%%%%%%%%%%%%%%%%%%%%%%
\paragraph{}

We begin with a review of the Newton-Cartan geometry and its torsional extensions. Unlike the relativistic case, where the fundamental geometric structure is the non-degenerate metric of a pseudo-Riemannian manifold, the Newton-Cartan geometry is described by a degenerate spatial metric $h^{\m\n}$ of rank-d and a temporal vielbein $\t_\m$ of rank-1, together with a connection $\G_{\m\n}^\r$ on an orientable manifold ${\mathcal M}$. Here, degeneracy imply that
\bea
h^{\m\n} \t_{\n} = 0 \,.
\label{Degeneracy}
\eea
In order to discuss the notions of parallel transport and geodesics we need to provide a suitable connection. In the relativistic case, the torsion-free  connection is uniquely fixed by the metric. As we will show in what follows, the connection in Newton-Cartan geometry is quite different: The uniqueness of the torsion-free compatible connection is lost and introducing torsion can break the invariance of the connection under Milne boosts. Moreover, the inclusion of non-metricity modifies the anti-symmetric part of the connection. As we will see, the degenerate nature of the Newton-Cartan geometry allows other geometric structures in addition to the ones we discussed above. Along the way, we will introduce necessary data to fix the connection uniquely. In the next two sections our focus will be understanding the connections with/without torsion which will be crucial obtaining the Newton-Cartan geometry from the gauging procedure.

\subsection{Torsionless Newton-Cartan Geometry}
\paragraph{}
Let us start our discussion with the torsionless Newton-Cartan geometry. With that in mind, we first impose that the connection $\G_{\m\n}^\r$ is symmetric and solve the metric compatibility conditions
\bea
\nabla_\m \t_\n = \partial_\m \t_{\n} - \G_{\m\n}^\r \t_{\r} &=& 0 \,,\nn\\
\nabla_\m h^{\n\r} =  \partial_{\m} h^{\n\r} + \G^\n_{\s\m} h^{\s\r}   + \G^\r_{\s\m} h^{\s\n} &=& 0\,,
\label{MetricCompatibility}
\eea
where the covariant derivative $\nabla$ is with respect to a connection $\G_{\m\n}^\r$. As the connection is symmetric, the antisymmetric part of the temporal metric compatibility condition implies
\bea
\t_\m = \partial_\m f \,,
\eea
for a scalar function $f(x^\m)$, which is chosen to be the absolute time $t$ so that the $f = const.$ simultaneity leaves foliate the spacetime. The temporal metric compatibility condition also fixes the temporal part of the connection as
\bea
\t_{\r} \G_{\m\n}^\r  = \partial_\m \t_{\n} \,.
\eea
Having determined the temporal part of the connection let us proceed with the spatial part. For that, we need to introduce two new tensors: The spatial inverse metric $h_{\m\n}$ and the temporal inverse vielbein $\t^\m$ which satisfies the following relations
\bea
h^{\m\s} h_{\n\s} = P^\m_\n = \d^\m_\n - \t^\m \t_\n \,, \qquad \t^\m \t_\m = 1 \,,\qquad h^{\m\n} \t_\n = 0 \,, \qquad h_{\m\n} \t^\n = 0 \,.
\label{InversionRelations}
\eea
Using the inverse quantities,  the most general symmetric connection compatible with the conditions (\ref{MetricCompatibility}) is given by \cite{Dautcourt}
\bea
\G_{\m\n}^\r = \t^\r \partial_\m \t_\n + \tfrac12 h^{\r\s} \Big( \partial_\n h_{\s\m} + \partial_\m h_{\s\n} - \partial_\s h_{\m\n} \Big) - h^{\r\s}\t_{(\m} F_{\n)\s}  \,,
\label{NewtonianConnection}
\eea
for an arbitrary 2-form $F_{\m\n}$. In order to make a contact with the covariant form of Newtonian gravity, i.e. Newton-Cartan gravity, the following conditions should be satisfied by the aptly named Newtonian connection:
 \begin{enumerate}
 \item {The geodesic equation based on $\G_{\m\n}^\r$ should give rise to the classical equation of motion of a massive particle
 	\bea
 	\frac{d^2 x^a (t)}{dt^2} + \frac{\partial \phi(x)}{d x^a} = 0 \,,
 	\label{Geodesic}
 	\eea
 where $x^a (t)$ are the spatial coordinates, $t$ is the absolute time and $\phi(x)$ is the Newtonian potential.}
 \item {The only non-vanishing component of the Riemann tensor
 	\bea
 	R^\m{}_{\n\r\s} (\G) = \partial_\r \G^\m_{\n\s} - \partial_\s\G^\m_{\n\r} +\G^\m_{\a\r} \G^\a_{\n\s}  - \G^\m_{\a\s} \G^\a_{\n\r} \,.
 	\label{RiemannTensor}
 	\eea
for the Newtonian connection (\ref{NewtonianConnection}) should give rise to the Poisson equation for the Newtonian potential,
 \bea
 \nabla^2 \phi  = 4 \pi G \rho \,,
 \label{Poisson}
 \eea
where $\r$ is the mass density.}
 \end{enumerate}
These two conditions can be satisfied given that the Riemann tensor (\ref{RiemannTensor}) satisfies the so-called Trautman \cite{Trautman} and Ehlers \cite{Ehlers} conditions
\bea
\label{Trautman}
&& h^{\s[\l} R^{\m]}{}_{(\n\r)\s} (\G) = 0 \,,\\
&& h^{\r\l} R^{\m}{}_{\n\r\s} R^{\n}{}_{\m\l\a} (\G)  = 0 \, \quad  \text{or} \quad   \t_{[\l} R^\m{}_{\n]\r\s} (\G)  = 0 \, \quad  \text{or} \quad   h^{\s[\l} R^{\m]}{}_{\n\r\s} (\G)  = 0 \,,
\label{Ehlers}
\eea
where the Trautman condition (\ref{Trautman}) further implies that for the connection to be Newtonian, $F_{\m\n}$ must be closed, i.e.
\bea
F_{\m\n} = 2 \partial_{[\m} m_{\n]} \,,
\label{KequalDM}
\eea
where $m_\m$ is a $\rm U(1)$ connection. With these conditions in hand, it is straightforward to show that the only non-vanishing component of the connection and the Riemann tensor are given by \cite{Andringa:2010it}
\bea
\G_{00}^a = \d^{ab} \partial_b \phi \,, \qquad  R^{a}{}_{0a0} (\G) = \nabla^2 \phi = 4 \pi G \r \,,
\eea
which satisfies the properties of a Newtonian connection. Thus, we conclude that the Newton-Cartan gravity is given by two degenerate metrics $h^{\m\n}$ and $\t_{\m}$ and a $\rm U(1)$ connection $m_\m$ equipped with the Trautman (\ref{Trautman}) and Ehlers (\ref{Ehlers}) conditions.

Before proceeding any further,  it is worth mentioning the Milne boost symmetry of the Newton-Cartan geometry and the invariant quantities in the presence of the $\rm U(1)$ connection. First of all, while the fundamental temporal and spatial metrics $\t_\m$ and $h^{\m\n}$ are uniquely defined, the inverse metrics $\t^\m$ and $h_{\m\n}$ in (\ref{InversionRelations}) are not unique, e.g. considering a 1-form $\p_\m$ we can define \cite{Jensen:2014aia} 
\bea
\t^{\prime \m} = \t^\m + h^{\m\n} \p_\n \,, \qquad h_{\m\n}^\prime = h_{\m\n} - (\t_\m  P_\n^\r + \t_\n  P_\m^\r)\p_\r + \t_\m \t_\n h^{\r\s} \p_\r \p_\s \,,
\label{redefinition1}
\eea
that still satisfies the inversion relations (\ref{InversionRelations}). These redefinitions are referred to as Milne boosts. The quantities built by using the connection $\G$ are covariant if the connection itself is invariant under the redefinition (\ref{redefinition1}). This would require the following Milne transformation property for the $\U(1)$ connection $m_\m$ \cite{Jensen:2014aia}
\bea
m_\m^\prime = m_\m - P_\m^\n \p_\n + \tfrac12 \t_\m h^{\n\r} \p_\n \p_\r \,,
\label{redefinition2}
\eea
in which case, the invariance of the connection $\G$ is satisfied given that its temporal part is symmetric \cite{Jensen:2014aia}. Thus, it is worth emphasizing that when introducing a temporal torsion, or torsion in general, one must be careful with the transformation of the connection under Milne boosts.

\subsection{Twistless Torsional Newton-Cartan Geometry}

\paragraph{}
In this subsection, we introduce a ``twistless torsion" to the Newton-Cartan geometry. As we will discuss in the detail below, the defining data of the twistless-torsional Newton-Cartan geometry (TTNC) is encoded in the following set of fields
%In this case, we consider a connection $\G_{\m\n}^\r$ on $\mathcal M$ that depends on these two metrics. 
\bea
(h^{\m\n}\,,  \t_{\m}\,,   b_\m \,,   M_\m ) \,,
\eea
where $b_\m$ and $M_\m$ are the necessary additional vector fields. To see the role of $b_\m$, we first consider the temporal component of the connection, which is fixed by the temporal metric compatibility condition (\ref{MetricCompatibility})
\bea
\t_{\r} \G_{\m\n}^\r =  \partial_{\m} \t_{\n } \,.
\eea
As a result, the time component of the torsion is fixed as
\bea
\t_\r \G_{[\m\n]}^\r =  \partial_{[\m} \t_{\n]} \,.
\eea
The  ``twistless torsion" condition is given by  \cite{Christensen:2013lma}
\bea
\t_{\l}\t_{[\r}  \G_{\m\n]}^\l =  \t_{[\r}  \partial_{\m} \t_{\n]} = 0 \,,
\eea
which indicates that the twistless torsional Newton-Cartan structure includes an additional Milne-invariant vector $b_\m$ by virtue of Frobenius theorem \cite{Bergshoeff:2014uea}
\bea
\partial_{[\m} \t_{\n]} = z b_{[\m} \t_{\n]} \,,
\label{Twistb}
\eea
where we introduced the coefficient $z$, the dynamical critical exponent, for later convenience. Next we determine the most general connection that is compatible with (\ref{MetricCompatibility}) by solving the compatibility condition for $h^{\m\n}$ \cite{Banerjee:2016laq}
\bea
\G_{\m\n}^\r &=& \t^\r \partial_\m \t_\n + \tfrac12 h^{\r\s} \Big( \partial_\n h_{\s\m} + \partial_\m h_{\s\n} - \partial_\s h_{\m\n} \Big) - h^{\r\s} \t_{(\m}F_{\n)\s} - K_{\m\n}{}^\r \,,
\label{TConn3}
\eea
where $F_{\m\n} = 2 \partial_{[\m} m_{\n]}$ and $K_{\m\n}{}^\r$ is the spatial contorsion tensor
\bea
\t_{\r} K_{\m\n}{}^\r = 0 \,, \qquad  h_{\r\s} (-K_{\m\n}{}^\s + K_{\n\m}{}^\s ) = h_{\r\s} T_{\m\n}{}^\s \,,
\eea
where $h_{\r\s} T_{\m\n}{}^\s$ defines the spatial part of the torsion. Recall that, previously the invariance of the connection under Milne boosts rely on the condition $\partial_{[\m} \t_{\n]}$ = 0.  In TTNC geometry we give up this condition, so the variation of the connection (\ref{TConn3}) is given by \cite{Jensen:2014aia}
\bea
\d_M \G_{\m\n}^\r &=&  h^{\r\s} \Big\{ (P_\s^\a \partial_{[\m} \t_{\n]} + P_\m^\a \partial_{[\s} \t_{\n]}   + P_\n^\a \partial_{[\s} \t_{\m]} ) \p_\a + \ft12 h^ {\a \b}\p_\a\p_\b (\t_\n \partial_{[\m} \t_{\s]}  + \t_\m \partial_{[\n} \t_{\s]}  ) \Big\}  \nn\\
&& - \d_M K_{\m\n}{}^\r  \,. \qquad\quad
\eea
Therefore, assuming that the contorsion tensor is $\rm U(1)$ invariant, $\d_{\rm U(1)} K_{\m\n}{}^\r = 0$, we can split the TTNC geometry into two cases depending on the Milne transformation of the contorsion tensor
\begin{enumerate}[c1.]
\item {If the spatial contorsion tensor is Milne-invariant $\d_M K_{\m\n}{}^\r = 0$, then we need to construct Milne-invariant inverse temporal and spatial metrics and a $\rm U(1)$ connection that still satisfies the inversion relations (\ref{InversionRelations}). The connection must be re-written based on the new inverse elements.}
\item {If the spatial contorsion transforms under the Milne transformation, then it must satisfy
\bea
\d_M K_{\m\n}{}^\r = h^{\r\s} \Big\{ (P_\s^\a \partial_{[\m} \t_{\n]} + P_\m^\a \partial_{[\s} \t_{\n]}   + P_\n^\a \partial_{[\s} \t_{\m]} ) \p_\a + \ft12 h^ {\a \b}\p_\a\p_\b (\t_\n \partial_{[\m} \t_{\s]}  + \t_\m \partial_{[\n} \t_{\s]}  ) \Big\} \,\,.
\label{MKmnr}
\eea}
\end{enumerate}
In the following, we will first consider these two cases separately and then show that one can transform between them by means of a linear transformation.

\subsection*{c1. Milne Invariant Spatial Contorsion Tensor}
\paragraph{}
We first consider a Milne invariant spatial contorsion tensor, $\d_{M} K_{\m\n}{}^\r = 0$, which includes a vanishing spatial contorsion as a special case. The only way to make the connection Milne invariant is to construct the connection in terms of Milne invariant objects. As given in (\ref{redefinition1}) and (\ref{redefinition2}), $\t^\m, h_{\m\n}$ and $m_\m$ are not invariant under Milne boosts. Milne invariance can be achieved by combining these quantities. However, as $m_\m$ is the $\rm U(1)$ connection, such combinations would fail the $\rm U(1)$ invariance. Therefore, we add a scalar field $\chi$ to the Newton-Cartan structure that transforms as shift under $\rm U(1)$ symmetry transformation
	\bea
	\d_{\rm U(1)} \chi = \s
	\label{Chi}
	\eea
	where $\s$ is the transformation parameter for the $\rm U(1)$ symmetry. We can now define a $\rm U(1)$ invariant vector field
	\bea
	M_\m = m_\m - \partial_\m \chi \,,
	\label{MmChi}
	\eea
	that still transforms as (\ref{redefinition2}) under the Milne boosts. Using this vector, we define a new, Milne invariant set of inverse metric fields \cite{Hartong:2015wxa,Hartong:2015zia}
	\bea
	\hat{\t}^\m &=& \t^\m + h^{\m\n} M_\n \,, \qquad \hat{h}_{\m\n} = h_{\m\n} -  \t_\m M_\n -  \t_\n M_\m + 2 \t_\m \t_\n \Phi  \,,
	\label{NewSet}
	\eea
where $\Phi$, the so-called Newton potential, is defined as
\bea
\Phi = \t^\s M_\s +  \tfrac{1}{2} h^{\r \s} M_\r M_\s \,.
\label{Potential}
\eea
%Note that we still have Milne invariant orthogonality property as
%	\bea
%	\hat{\t}^\m\hat{h}_{\m\n} &=& 0 \label{ort}
%\eea
This new set of fields $(\hat{\t}^\m, \hat{h}_{\m\n}, M_\m)$  satisfy the inversion relations (\ref{InversionRelations}), and we give the connection that solves the metric compatibility condition (\ref{MetricCompatibility}) as
\bea
\widehat{\G}_{\m\n}^\r &=& \hat\t^\r \partial_\m \t_\n + \tfrac12 {h}^{\r\s} \Big( \partial_\n \hat{h}_{\s\m} + \partial_\m \hat h_{\s\n} - \partial_\s \hat h_{\m\n} \Big) + h^{\r\s} \t_{\m} \t_{\n}  \partial_{\s} \Phi - K_{\m\n}{}^\r  \,,
\label{TConn4}
\eea
where the spatial contorsion tensor is now invariant under Milne boosts and $\rm U(1)$ transformations. Here, we also introduce a hatted-connection, $\widehat{\G}$, in order to emphasize that this connection is constructed by use of hatted inverse metrics. Note that the penultimate term in the connection (\ref{TConn4}) corresponds to a special choice of the arbitrary function $F_{\m\n}$, which we made by demanding that when both the spatial and the temporal torsion vanishes we recover the standard Milne invariant Newton-Cartan connection (\ref{NewtonianConnection}), i.e.
\begin{figure}
\centering
\begin{tikzpicture}
%\draw[step=1cm, gray, very thin] (0,0) grid (10,10);
\node at (5,8.5) {\underline{TTNC}};
\node at (5,8) {$(h^{\m\n},\t_\m, b_\m, M_\m)$};
%\draw (3,7) -- (7,7) -- (7,8.5) -- (3,8.5) -- (3,7) ;
\node at (2,6) {\underline{$\d_M K_{\m\n}{}^\r = 0$}};
\node at (2,5.4) {$(\hat{h}_{\m\n},\hat\t^\m)$};
\node at (8,6) {\underline{$\d_M K_{\m\n}{}^\r \neq  0$}};
\node at (8,5.4) {$({h}_{\m\n},\t^\m)$};
\draw [thick, ->]  (3.3,6.1) -- (6.7,6.1) ;
\node at (5,6.1) [above] {Eq.(\ref{NewSet})} ;
\draw [thick, <-]  (3.3,5.7) -- (6.7,5.7) ;
\node at (5,5.7) [below] {Eq.(\ref{OldNew})} ;
\draw [thick, ->]  (4,7.7) -- (2,6.4) ;
\draw [thick, ->]  (6,7.7) -- (8,6.4) ;
\end{tikzpicture}
\caption{The schematic relation between the c1 and c2 cases. The defining data of TTNC geometry is given by $(h^{\m\n},\t_\m, b_\m, M_\m)$ and the inverse vielbein and the spatial metric are chosen depending on the Milne transformation of the contorsion tensor. Different choices are related to each other by means of linear maps (\ref{NewSet}) and (\ref{OldNew}). }
\label{TTNC}
\end{figure}
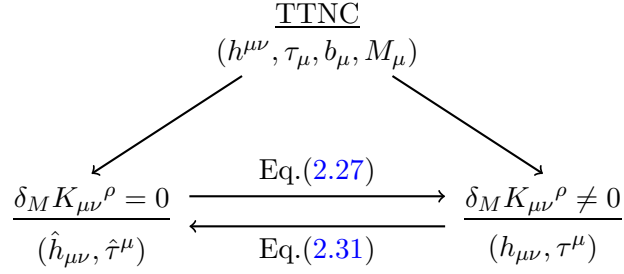
\bea
\partial_{[\m} \t_{\n]} = 0 \, \quad \text{and} \quad K_{\m\n}{}^\r = 0  \quad \Rightarrow \quad \widehat{\G}_{\m\n}^\r = {\G}_{\m\n}^\r \,.
\eea

\subsection*{c2. Non-Invariant Spatial Contorsion Tensor}
\paragraph{}

The second choice is to work with a non-invariant contorsion tensor, then its Milne transformation is given by (\ref{MKmnr}). In this case, we work with the original set of inverse metrics (\ref{InversionRelations}) and the connection is given by (\ref{TConn3}). Note that the connection includes $m_\m$ via its field strength, therefore the use of $M_\m$ leaves the connection unchanged.

Although we investigated the twistless torsional case in two separate cases, they are not independent from each other since the new set of inverse elements $(\hat{\t}^\m, \hat{h}_{\m\n})$ and the original set $({\t}^\m, {h}_{\m\n})$ can be transformed to each other by means of a linear transformation (\ref{NewSet}), see Fig \ref{TTNC}. To see that, we can consider a connection $\G_{\m\n}^\r$ with a non-invariant contorsion tensor $K_{\m\n}{}^\r$ given by (\ref{TConn3}) and replace $({\t}^\m, {h}_{\m\n})$ with $(\hat{\t}^\m, \hat{h}_{\m\n})$ via
\bea
{\t}^\m &=& \hat{\t}^\m - h^{\m\n} M_\n \,, \quad {h}_{\m\n} = \hat{h}_{\m\n} +   \t_\m M_\n +  \t_\n M_\m -  2 \t_\m \t_\n \Phi  \,.
\label{OldNew}
\eea
Upon this replacement, we obtain a connection with a Milne-invariant contorsion tensor
\bea
\G_{\m\n}^\r = \widehat{\G}_{\m\n}^\r - K^\prime_{\m\n}{}^\r \,,
\label{GGbar}
\eea
where the Milne-invariant contorsion, $K^\prime_{\m\n}{}^\r$, is related to the non-invariant contorsion tensor $K_{\m\n}{}^\r$ via
\bea
K^\prime_{\m\n}{}^\r = K_{\m\n}{}^\r - h^{\r\s} (M_{\s} \partial_{[\m} \t_{\n]} + M_{\n} \partial_{[\s} \t_{\m]} + M_{\m} \partial_{[\s} \t_{\n]}  - 2 \Phi \t_\m \partial_{[\s} \t_{\n]} - 2 \Phi \t_\n \partial_{[\s} \t_{\m]}  )  \,.
\label{RelCont}
\eea
With this result in hand, it is straightforward to generalize our discussion to an arbitrary torsion.
If we introduce an arbitrary torsion to the Newton-Cartan geometry, the temporal component of the torsion is not subject to any constraint, and is again fixed by the temporal metric compatibility (\ref{MetricCompatibility}). Furthermore, the most general connection is still given by (\ref{TConn3}) and the Milne invariance of the connection is again achieved by following the previous discussion for the TTNC.

%%%%%%%%%%%%%%%%%%%%%%%%%%%%%%%%%%%%%%%%%%%%%%%%%%%%%%%%%%%%%%%%%
\section{Non-Relativistic Scale Symmetry and Newton-Cartan Geometry} {\label{Section3}}
%%%%%%%%%%%%%%%%%%%%%%%%%%%%%%%%%%%%%%%%%%%%%%%%%%%%%%%%%%%%%%%%%
 \paragraph{}
 In this section, we introduce the non-relativistic analogue of the scale symmetry to Newton-Cartan geometry. The defining property of the scale symmetry is via breaking of the compatibility condition (\ref{MetricCompatibility}) by a particular non-metricity tensor
 \bea
 \nabla_\m \t_\n =  z b_\m \t_\n \,, \qquad \nabla_\m h^{\n\r} = - 2 b_\m h^{\n\r} \,,
 \label{NonMetricity}
 \eea
 which is preserved by the following transformations
 \bea
 \t_{\m} \rightarrow e^{z \L_D (x)} \t_\m \,, \qquad h^{\m\n} \rightarrow e^{-2\L_D (x)} h^{\m\n} \,, \qquad b_\m  \rightarrow b_\m + \partial_\m \L_D (x)  \,, \qquad \G_{\m\n}^\r \rightarrow \G_{\m\n}^\r  \,,
 \label{Scalings}
 \eea
where vector field $b_\m$ is the gauge field for the scale transformations and is Milne invariant
\bea
\d_M b_\m = 0 \,.
\eea
Here, we purposefully represent the gauge field of the scale transformations with the same field that we used to define the twistless torsional condition (\ref{Twistb}) to keep the number of fields minimum. 
Furthermore, it is important to note that in general the spatial metric and the temporal vielbein have different scaling dimensions which only coincide for $z=1$. This is the reminiscent of the Schr\"odinger symmetries in $d$ spatial dimensions with $z$ critical exponent that transform the time $(t)$ and space $(\bold{x})$ coordinates under dilatation with a rigid dilatation parameter $\l$ as follows
\bea
\bold{x} \, \rightarrow \l \, \bold{x}\,, \qquad t \rightarrow \l^z \, t \,.
\eea
In order to solve the connection in terms of the Newton-Cartan variables, we first consider the scale covariant temporal compatibility condition (\ref{NonMetricity}) which fixes the temporal part of the connection
 \bea
 \t_\r \G_{\m\n}^\r = \partial_{\mu} \t_{\n} - z b_{\m} \t_{\n} \,.
 \label{TemporalGamma}
 \eea
 At this point, some clarifications are in order
 \begin{itemize}
 \item {Unlike the relativistic scenarios, the inclusion of the non-metricity modifies the anti-symmetric part of the connection
 \bea
 \t_\r \G_{[\m\n]}^\r = \partial_{[\mu} \t_{\n]} - z b_{[\m} \t_{\n]} \,.
\eea
Thus, when the twistless condition is imposed
\bea
\partial_{[\mu} \t_{\n]}  = z b_{[\m} \t_{\n]} \,,
\eea
the anti-symmetric part of the temporal part of the connection, thereby the temporal torsion, vanishes.
}
 \item {In principle, one might think that a twistless torsion can be introduced if the twistless condition is imposed by means of another vector field $A_\m$ such that
 \bea
\partial_{[\mu} \t_{\n]}  = A_{[\m} \t_{\n]} \,.
\eea
However, in this case, the scaling transformation of $\t_{\m}$ forces us to set $A_\m = z b_\m$.}
\item {When the torsion is arbitrary we only impose 
	\bea
	\partial_{[\m} \t_{\n]} \neq z b_{[\m} \t_{\n]}\,,
	\eea
	 and the temporal part of the connection reads (\ref{TemporalGamma}). }
 \end{itemize}
In the following, we construct the connection in terms of the Milne invariant set of inverse fields $(\hat{\t}^\m, \hat{h}_{\m\n}, M_\m)$.  The scaling properties of the Milne invariant set are given by
\bea
\hat{\t}^\m \rightarrow e^{-z \L_D (x)} \hat{\t}^\m \,, \quad \hat{h}_{\m\n} \rightarrow e^{2\L_D (x)} \hat{h}_{\m\n} \,, \quad M_\m \rightarrow e^{-(z-2)\L_D (x)} M_\m \,.
\eea
Solving the  scale covariant compatibility conditions, the connection reads
\bea
\hat{\G}_{\m\n}^\r &=& \hat\t^\r \cD_\m \t_\n + \tfrac12 {h}^{\r\s} \Big( \cD_\n \hat{h}_{\s\m} + \cD_\m \hat h_{\s\n} - \cD_\s \hat h_{\m\n} \Big) + h^{\r\s} \t_{\m} \t_{\n}  \cD_{\s} \Phi - K_{\m\n}{}^\r  \,,
\label{SConn}
\eea
where $K_{\m\n}{}^\r$ is a scale and Milne invariant spatial contorsion tensor
\bea
\d_D K_{\m\n}{}^\r = \d_M K_{\m\n}{}^\r = 0 \,,
\eea
 and the scale-covariant derivatives are defined as
 \bea
 \cD_{\m} \t_{\n} = \partial_{\m} \t_{\n} - z b_{\m} \t_{\n}\,, \quad \cD_{\m} \hat{h}_{\n\r} = \partial_{\m}  \hat{h}_{\n\r} - 2 b_\m \hat{h}_{\n\r}\,, \quad \cD_\m \Phi = \partial_{\m} \Phi + (2z - 2) b_\m \Phi \,.
 \label{ScaleCovariant}
 \eea
 Note that in the presence of the scale transformations, the definition of $M_\m$ in terms of $m_\m$ and $\chi$ as given in (\ref{MmChi}) needs to be modified with a $b_\m$ dependent term as \cite{Bergshoeff:2014uea}
 \bea
 M_\m = m_\m - \partial_\m \chi - (z-2) b_\m \chi \,,
 \label{Mmubmu}
 \eea
 where the scale transformations of $m_\m$ and $\chi$ are given by
 \bea
 m_\m \rightarrow e^{(2-z) \L_D} m_\m \,, \quad \chi \rightarrow e^{(2-z) \L_D} \chi \,.
 \eea
 The definition of $M_\m$ as given in (\ref{Mmubmu}) also implies that the $\U(1)$ transformation of $m_\m$ must be modified with a $b_\m$ dependent term as follows
 \bea
 m_\m \rightarrow  m_\m + \partial_{\m} \s + (z-2) \s b_\m \,.
 \eea
 Based on the scale-invariant connection (\ref{SConn}) in hand, it is now straightforward to write a scale-invariant Riemann tensor \cite{Mitra:2015twa} as well as a $(d+1)$ Milne-invariant tensor $g_{\m\n}$   \cite{Jensen:2014aia}
 \bea
 g_{\m\n} = \hat{h}_{\m\n} + \t_\m \t_\n \,, \qquad  g^{\m\n} = {h}^{\m\n} + \hat\t^\m \hat\t^\n \,,
 \label{gmn}
 \eea
 to be used to define a volume form on ${\mathcal M}$.
% as
% \bea
% {R}^\m{}_{\n\r\s} ({\G}) \equiv  \partial_\r \G^\m_{\n\s} - \partial_\s\G^\m_{\n\r} +\G^\m_{\a\r} \G^\a_{\n\s}  - \G^\m_{\a\s} \G^\a_{\n\r} \,,
% \eea
With these results in hand, one can also construct non-relativistic scale invariant gravity actions or scalar field equations by utilizing a real scalar field with a Weyl weight $\o$ that transforms as
\bea
\d \phi = \o \L_D \phi \,.
\label{Phi}
\eea
 Finally, it is important to note that although we worked with a particular set of inverse fields, we expect that it is always possible to switch between different sets as described in Fig \ref{TTNC}.
% Therefore, if a contorsion tensor with a Milne transformation is under consideration, we give the connection as
% \bea
% \G_{\m\n}^\r = \widetilde{\G}_{\m\n}^\r - K^\prime_{\m\n}{}^\r \,,
% \label{GGbar2}
% \eea
 However, due to scale invariance of the connection, the relation between the contorsion tensors (\ref{RelCont}) is modified as
 \bea
 K^\prime_{\m\n}{}^\r = K_{\m\n}{}^\r -  h^{\r\s} (M_{\s} \cD_{[\m} \t_{\n]} + M_{\n} \cD_{[\s} \t_{\m]} + M_{\m} \cD_{[\s} \t_{\n]}  - 2 \Phi \t_\m \cD_{[\s} \t_{\n]} - 2 \Phi \t_\n \cD_{[\s} \t_{\m]}  )  \,.
 \eea
This indicates that as $\cD_{[\m} \t_{\n]} = 0$ for the twistless torsional case, it is not possible to introduce a contorsion with a non-trivial Milne transformation - the Galilean invariance of the connection is maintained by the non-metricity property of the temporal vielbein $\t_{\m}$.

\begin{table}[]
\centering
\begin{tabular}{|l|c|c|c|c|}
\hline
Field            & Milne                                                 & Scaling                & $\U(1)$                          & Special Conformal \\ \hline
$\t_\m$          & 0                                                     & $z \L_D \t_\m$         & 0                                & 0                 \\ \hline
$h^{\m\n}$       & 0                                                     & $-2\L_D h^{\m\n}$      & 0                                & 0                 \\ \hline
$\hat{\t}^\m$    & 0                                                     & $-z\L_D \hat{\t}^\m$   & 0                                & 0                 \\ \hline
$\hat{h}_{\m\n}$ & 0                                                     & $2\L_D \hat{h}_{\m\n}$ & 0                                & 0                 \\ \hline
$b_\m$           & 0                                                     & $\partial_\m \L_D$     & 0                                & $\t_\m \L_{K}$    \\ \hline
$m_\m$           & $- P_{\m}^\n \p_\n +\frac12 \t_\m h^{\r\s} \p_\r \p_\s$ & $(2-z)\L_D m_\m$       & $\partial_\m \s + (z-2) b_\m \s$ & 0                 \\ \hline
$M_\m$           & $-P_{\m}^\n \p_\n +\frac12 \t_\m h^{\r\s} \p_\r \p_\s$ & $(2-z)\L_D M_\m$       & 0                                & 0                 \\ \hline
$\chi$           & 0                                                     & $(2-z) \L_D \chi $     & $\s$                             & 0                 \\ \hline
$\phi $       & 0                                                   & $\o \L_D \phi $     &  0                         & 0               \\ \hline
\end{tabular}
\vspace{0.3cm}
\caption{Transformation rules for the fields in scale and Schr\"odinger extended Newton-Cartan gravity. When the Newton-Cartan symmetries are extended by only a scale symmetry, the appearance the critical exponent $z$ is allowed. When scale symmetry is enhanced to Schr\"odinger symmetry, we set $z=2$.}
\label{T1}
\end{table}
%
%\subsection{Schr\"odinger Symmetry}
%\paragraph{}
Having extended the Newton-Cartan geometry with a scale symmetry, let us consider the case when $z = 2$, the value for which the non-relativistic scale symmetry can be enhanced to the Schr\"odinger symmetry, i.e. the non-relativistic analogue of the conformal symmetry, by introducing a special conformal transformation. This is done by imposing the $z=2$ scale covariant compatibility conditions (\ref{NonMetricity})
\bea
 \nabla_\m \t_\n =  2 b_\m \t_\n \,, \qquad \nabla_\m h^{\n\r} = - 2 b_\m h^{\n\r} \,,
\eea
which is preserved by the non-relativistic special conformal transformation
\bea
b_\m \rightarrow b_\m +  \t_\m \L_{K}(x) \,.
\label{TransformB}
\eea
Since the compatibility conditions are not modified, the connection is still given by (\ref{SConn}), which  transforms non-trivially under special conformal transformations due to appearance of $b_\m$ in its definition
\bea
\G_{\m\n}^\r \rightarrow  \G_{\m\n}^\r  - \t_\m \d_{\n}^\r \L_K -\t_\n \d_{\m}^\r \L_K \,.
\label{GammaB}
\eea
Therefore, we conclude that as the temporal component of $b_\m$ is the only field that transforms non-trivially under the special conformal transformation, see Table \ref{T1},  a Schr\"odinger invariant gravity means that a $z=2$ scale invariant gravity that does not contain any $\t^\m b_\m$ term.

%%%%%%%%%%%%%%%%%%%%%%%%%%%%%%%%%%%%%%%%%%%%%%%%%%%%%%%%%%%%%%%%%
\section{$z=2$ Scale and Schr\"odinger Symmetry}{\label{Sec3}}
%%%%%%%%%%%%%%%%%%%%%%%%%%%%%%%%%%%%%%%%%%%%%%%%%%%%%%%%%%%%%%%%%
\paragraph{}
In the previous section, we approach to the Newton-Cartan gravity as well as its torsional, scale and Schr\"odinger extensions from a geometric perspective. In particular, we put a distinction between the realization of local scale and Schr\"odinger symmetry in non-relativistic gravity, and stated that a Schr\"odinger invariant gravity is a $z=2$ scale invariant gravity that does not contain any $\t^\m b_\m$ term.  We now want to make this distinction explicit by constructing scale invariant Newton-Cartan models and show that the scale symmetry is enhanced to the non-relativistic conformal symmetry when particular models are combined to annihilate all  $\t^\m b_\m$ terms. Such constructions are most simply done by gauging the relevant spacetime symmetries. Thus, we dedicate this section to introduce the Bargmann algebra and its scale and Schr\"odinger extensions.

%In relativistic case, considering gravity as a gauge theory is based on the gauging of the Poincar\'e algebra, and scale and conformal invariant gravity models can simply be obtained by extending the Poincar\'e algebra with the corresponding generators. Here, our approach to the construction of Newton-Cartan geometry and its torsional extensions is similar: the gauging of the Bargmann algebra gives rise to the Newton-Cartan geometry while its scale and conformal extensions, which give rise to TTNC, are obtained by extending the Bargmann algebra with a dilatation and special conformal symmetry generator.

\subsection{Bargmann Algebra}
%%%%%%%%%%%%%%%%%%%%%%%%%%%%%%%%%%%%%%%%%%%%%%%%%%%%%%%%%%%%%%%%%
\paragraph{}
To set the stage, we first review the basics of the Bargmann algebra, the central extension of the Galilean algebra that is generated by the time translations $H$, space translations $P_a$, Galilean boosts $G_a$, and the spatial rotations $J_{ab}$. Here, Galilean boosts represent the infinitesimal realization of the Milne boosts that we discussed in the previous section. Considering Newton-Cartan gravity as a gauge theory is based on the Bargmann algebra with the following generators
\bea
\{  H\,, \quad P_a \,, \quad J_{ab} \,, \quad G_a \,, \quad N \} \,,
\eea
where $N$ for central charge transformations. The commutation relations between the generators of the Bargmann algebra is given by \cite{Andringa:2010it}
\begin{eqnarray}\label{Bargmann}
&&  [H,G_a] = P_a\,,\quad  [P_a,G_b] = \delta_{ab}N\,,\quad [J_{ab}, P_c] = 2\delta_{c[a}P_{b]}\,, \nn\\
&&  [J_{ab}, G_c] = 2\delta_{c[a}G_{b]}\,, \quad  [J_{ab}, J_{cd}] = 4\delta_{[a[c}\,J_{b]d]}\,.
\end{eqnarray}
The gauge fields corresponding to these generators are
\bea
h_\m{}^A &=& \{ \t_\m \,, \quad   e_\m{}^a \,, \quad \o_{\m}{}^{ab} \,, \quad \o_{\m}{}^a \,, \quad m_\m  \} \,,
\label{GaugeFields}
\eea
where the Latin indices $a,b,c,\ldots$ refer to the spatial local Galilean frame while the Greek indices $\m,\n\ldots$ refer to the coordinate frame and labels all spacetime coordinates, $x \equiv (t, {x^i})$. The transformations are generated by operators according to
\bea
\d &=& \x H + \x^a P_a + \tfrac{1}{2}\l^{ab} J_{ab} + \l^a G_a + \s N \,,
\eea
where $\x, \x^a,\l^a, \l^{ab}$ and  $\s $ are the parameters for the time translations,  space translations, Galilean boosts, spatial rotations and central charge transformations in the respective order. Using the structure constants of the Bargmann algebra \cite{Andringa:2010it} and the standard rules
\bea
\d h_\m{}^A &=& \partial_\m \e^A + f_{BC}{}^A h_\m{}^B \e^C \,, \nn\\
R_{\m\n}{}^A &=& 2 \partial_{[\m} h_{\n]}{}^A + f_{BC}{}^A h_\m{}^B h_\n{}^C \,, \label{curv}
\eea
we give the transformation rules for the gauge fields as \cite{Andringa:2010it}
\bea
\d \t_\m &=& \partial_\m \xi \,,\nn\\
\d e_\m{}^a &=& \partial_\m \xi^a  - \o_\m{}^{ab} \x_b  + \l^a{}_b e_\m{}^b + \l^a \t_\m - \o_\m{}^a \x \,,\nn\\
\d \o_\m{}^{ab} &=& \partial_\m \l^{ab} + 2\l^{c[a} \o_\m{}^{b]}{}_c \,,\nn\\
\d \o_\m{}^a &=& \partial_\m \l^a - \o_{\m}{}^{ab} \l_b + \l^{a}{}_{b}\,  \o_{\m}{}^{b}  \,,\nn\\
\d m_\m &=& \partial_\m \s - \x^a \o_{\m a} + \l^a e_{\m a} \,.
\eea
and the corresponding curvatures are given by
\bea
R_{\m\n}(H) &=& 2 \partial_{[\m} \t_{\n]} \,,\nn\\
R_{\m\n}{}^a (P) &=&  2 \partial_{[\m} e_{\n]}{}^a - 2 \o_{[\m}{}^{ab} e_{\n]b} - 2 \o_{[\m}{}^a \t_{\n]} \,,\nn\\
R_{\m\n}{}^{ab} (J) &=& 2 \partial_{[\m} \o_{\n]}{}^{ab} - 2 \o_{[\m}{}^{c[a} \o_{\n]}{}^{b]}{}_c \,,\nn\\
R_{\m\n}{}^a (G) &=& 2 \partial_{[\m} \o_{\n]}{}^a + 2 \o_{[\m}{}^{b} \o_{\n]}{}^a{}_b  \,,\nn\\
R_{\m\n} (N) &=& 2 \partial_{[\m} m_{\n]}  - 2\o_{[\m}{}^a e_{\n]a} \,.
\eea
In order to leave the $\t_\m, e_\m{}^a$ and $m_\m$ as the only independent fields, we impose the following constraints \cite{Andringa:2010it}
\bea
R_{\m\n}(H) = 0 \,, \qquad  R_{\m\n}{}^a (P) = 0 \,, \qquad R_{\m\n}(N) = 0 \,,
\label{CC1}
\eea
and the Bianchi identity on $R_{\m\n}{}^a (P)$ and $R_{\m\n}(N)$ gives rise to the following relations between curvatures
\bea
e_{[\m}{}^b R_{\n\r]}{}^{a}_{\  \ b} (J)  + \t_{[\m}R_{\n\r]}{}^a (G) = 0 \,,\qquad e_{[\m}{}^a R_{\n\r]a} (G) = 0 \,.
\label{B1}
\eea
Here, first two equations in (\ref{CC1}) correspond to the non-relativistic version of the torsionless  structure equations (\ref{MetricCompatibility}). Furthermore, as the first equation reads $\partial_{[\m} \t_{\n]} = 0$, we may take
\bea
\t_{\m} = \partial_\m t \,.
\eea
In order to solve the last two equations in (\ref{CC1}) to obtain composite expressions for $\o_{\m}{}^{ab}$ and $\o_{\m}{}^a$, we introduce two new fields: The inverse temporal vielbein $\t^\m$ and the inverse spatial vielbein $e^\m{}_{a}$ with the following properties
\bea
 \t_{\m} e^{\m}{}_a = 0\,, \quad \t^\m e_\m{}^a = 0 \,, \quad \t^\m \t_{\m} = 1 \,,\quad  e_\m{}^a e^\m{}_b = \d^a_b \,, \quad e_\m{}^a e^\n{}_a = \d_\m^\n - \t_{\m} \t^\n \,.
\eea
Note that these properties are invariant under the Galilean boost transformations given that $\t^\m$ and $e^\m{}_a$ transforms under Galilean boost transformations as
\bea
\d_G \t^\m = - \l^a e^\m{}_a \,, \qquad \d_G e^\m{}_a = 0 \,.
\label{InverseGal}
\eea
Furthermore, any tensor $T_{\m}$ can be decomposed into its spatial and temporal part using the inverse temporal and spatial vielbein
\bea
T_\m = \t_{\m} T_0  + e_\m{}^a T_a \,, \quad \text{such that} \quad T_0 = \t^\m T_\m \,, \quad \text{and} \quad  T_a = e^\m{}_a T_\m \,.
\eea
Using these inverse fields, we give the composite expressions for $\o_{\m}{}^{ab}$ and $\o_{\m}{}^a$ as
\bea
\o_\m{}^{ab} &=& -2 e^{\n [a} \partial_{[\m} e_{\n]}{}^{b]} + e^{\n a} e^{\r b} e_{\m c} \partial_{[\n} e_{\r]}{}^c
- \t_\m e^{\n a} e^{\r b} \partial_{[\n} m_{\r]} \,,\nn\\
\o_\m{}^a  &=& \t^\n \partial_{[\m} e_{\n]}{}^a + e^{\n a} \t^\r e_{\m b} \partial_{[\n} e_{\r]}{}^b
+ e^{\n a} \partial_{[\m} m_{\n]} + \t_\m \t^\n e^{\r a} \partial_{[\n} m_{\r]} \,.
\label{Df1}
\eea
In order to make contact with Newton-Cartan geometry, we first determine $\t_{\m}$ as the temporal metric and define the spatial metric $h^{\m\n}$ in terms of the inverse spatial vielbein $e^\m{}_a$ as
\bea
h^{\m\n} = e^\m{}_a \, e^{\n}{}_b \, \d^{ab} \,.
\label{InvMetricE}
\eea
Then, the connection can be determined by the spatial and temporal vielbein postulates
\bea
 \partial_\m e_\n{}^a - \o_{\m}{}^{ab} e_{\n b} - \o_\m{}^a \t_\n - \G_{\m\n}^\r e_{\r}{}^a = 0 \,, \qquad  \partial_\m \t_\n - \G_{\m\n}^\r \t_\r  = 0 \,.
\eea
These two equations fix the connection as
\bea
\G_{\m\n}^\r &=& \t^\r \partial_{\m} \t_{\n} + e^\r{}_a \Big( \partial_{\m} e_\n{}^a - \o_{\m}{}^{ab} e_{\n b} - \o_{\m}{}^a \t_{\n} \Big) \,,
\eea
which is the same symmetric connection that we obtained by imposing the torsionless metric compatibility condition (\ref{NewtonianConnection}) upon using the definition of the inverse spatial metric (\ref{InvMetricE}) and the composite expressions (\ref{Df1}). The Riemann tensor corresponding to this connection is given in terms of $R_{\m\n}{}^{ab} (J)$ and $R_{\m\n}{}^a (G)$ as \cite{Andringa:2010it}
\bea
R^\r{}_{\m\n\s} (\G)= - e^\r{}_a \Big( \t_\m R_{\n\s}{}^a (G) + e_{\m b} R_{\n\s}{}^{ab}(J)  \Big) .
\eea
In order for this connection to be the Newtonian connection, we must satisfy the  Trautman (\ref{Trautman}) and Ehlers conditions (\ref{Ehlers}), which can be achieved by the following curvature constraint \cite{Afshar:2015aku}
\bea
e^\n{}_a R_{\m\n}{}^{ab} (J) = 0 \,.
\eea
This constraint implies that the only non-vanishing component of $R_{\m\n}{}^a (G)$ is related to the only non-zero component of the Riemann tensor \cite{Andringa:2010it}
\bea
\t^\m e^{\n (a}R_{\m\n}{}^{b)} (G) = \d^{c(a} R^{b)}{}_{0c0} (\G) \,,
\eea
Therefore we obtain the desired geodesic equation (\ref{Geodesic}) and the Poisson equation (\ref{Poisson}) by means of the gauging of the Bargmann algebra.

\subsection{z=2 Non-Relativistic Scale Invariance}
\paragraph{}
When an additional local scale invariance is demanded, we can improve the Bargmann group with the scale symmetry generator $D$
\bea
\{ H\,, \quad P_a \,, \quad J_{ab} \,, \quad G_a \,, \quad N\,, \quad D\} \,.
\eea
which introduces an additional gauge field for scale transformations to (\ref{GaugeFields}), which we represent by $b_\m$. The commutation relations between the generators of the $z=2$ scale extended Bargmann algebra is given by
\begin{eqnarray}\label{z=2SBargmann}
&&[D,P_a] = -P_a\,,\qquad   [D,H] = -2H\,,\qquad  [H,G_a] = P_a\,,\nonumber\\[.2truecm]
&& [P_a,G_b] = \delta_{ab}N\,, \qquad [D,G_a] = G_a\,, \qquad [J_{ab}, P_c] = 2\delta_{c[a}P_{b]}\,, \nn\\
&& [J_{ab}, G_c] = 2\delta_{c[a}G_{b]}\,, \qquad  [J_{ab}, J_{cd}] = 4\delta_{[a[c}\,J_{b]d]}\,.
\end{eqnarray}
As mentioned in the previous section, the addition of scale invariance can be made while keeping the dynamical critical exponent $z$ arbitrary. Here, we are interested in comparing the scale invariance with the Schr\"odinger invariance, thus we set $z=2$ for the rest of this section.

Using the structure constants for the scale-extended Bargmann group \cite{Bergshoeff:2014uea}, we give the transformation rules for the gauge fields as
\bea
\d \t_\m &=& \partial_\m \xi - 2\xi b_\mu +  2 \L_D \t_\m\,,\nn\\
\d e_\m{}^a &=& \partial_\m \xi^a  - \o_\m{}^{ab} \x_b  - b_\m \x^a + \l^a{}_b e_\m{}^b + \l^a \t_\m - \o_\m{}^a \x + \L_D e_\m{}^a \,,\nn\\
\d \o_\m{}^{ab} &=& \partial_\m \l^{ab} + 2\l^{c[a} \o_\m{}^{b]}{}_c \,,\nn\\
\d \o_\m{}^a &=& \partial_\m \l^a - \o_{\m}{}^{ab} \l_b + \l^{a}{}_{b}\,  \o_{\m}{}^{b} +\l^a b_\m -\L_D \o_\m^{\ a} \,,\nn\\
\d m_\m &=& \partial_\m \s - \x^a \o_{\m a} + \l^a e_{\m a} \,, \nn \\
\d b_\m &=& \partial_\m \L_D \,,
\label{SITrans}
\eea
and the corresponding curvatures are given by
\bea
R_{\m\n}(H) &=& 2 \partial_{[\m} \t_{\n]} - 4 b_{[\m} \t_{\n]} \,,\nn\\
R_{\m\n}{}^a (P) &=&  2 \partial_{[\m} e_{\n]}{}^a - 2 \o_{[\m}{}^{ab} e_{\n]b} - 2 \o_{[\m}{}^a \t_{\n]} - 2 b_{[\m} e_{\n]}{}^a \,,\nn\\
R_{\m\n}{}^{ab} (J) &=& 2 \partial_{[\m} \o_{\n]}{}^{ab} - 2 \o_{[\m}{}^{c[a} \o_{\n]}{}^{b]}{}_c \,,\nn\\
R_{\m\n}{}^a (G) &=& 2 \partial_{[\m} \o_{\n]}{}^a + 2 \o_{[\m}{}^{b} \o_{\n]}{}^a{}_b - 2 \o_{[\m}{}^{a} b_{\n]} \,,\nn\\
R_{\m\n} (D) &=& 2 \partial_{[\m} b_{\n]}\,,\nn\\
R_{\m\n} (N) &=& 2 \partial_{[\m} m_{\n]} - 2\o_{[\m}{}^a e_{\n]a} \,.
\label{ScaleCurv}
\eea
Note that in the presence of the dilatation, the $G$ and $D$ transformation rules for the inverse vielbein and the inverse temporal vielbein are given by
\bea
\d \t^\m =  - 2 \L_D \t^\m - \l^a e^\m{}_a \,, \qquad  \d e^\m{}_a =    - \L_D e^\m{}_a \,.
\eea

In order to make contact to the scale-invariant generalization of the Newton-Cartan geometry, which we established in Section \ref{Section3}, we want to solve $\o_\m{}^{ab}$ and $\o_{\m}{}^a$ in terms of the other fields to leave $\t_{\m}, e_\m{}^a, m_\m$ and $b_\m$  as the set of fields that characterizes the scale-invariant Newton-Cartan geometry. This can be achieved by the following set of constraints
\bea
R_{\m\n}(H) = 0 \,, \qquad  R_{\m\n}{}^a (P) = 0 \,, \qquad R_{\m\n}(N) = 0 \,,
\label{CC2}
\eea
which results to the following further constraints by Bianchi identities
\bea
e_{[\m}{}^b R_{\n\r]}{}^{a}{}_{b} (J) + e_{[\m}{}^aR_{\n\r]}{} (D) + \t_{[\m}R_{\n\r]}{}^a (G) = 0 \,, \quad
\t_{[\m}R_{\n\r]}{} (D) = 0  \,,\quad e_{[\m}{}^a R_{\n\r]a} (G) = 0 \,. \label{B2}
\eea
The first constraint implies that the twistless condition is satisfied, thereby the torsion vanishes 
\bea
R_{\m\n} (H)  = 0 \qquad \Rightarrow \qquad \partial_{[\mu} \t_{\n]}  = 2 b_{[\m} \t_{\n]} \,.
\eea
Furthermore, the last two constraints in (\ref{CC2}) gives rise to the solution of $\o_{\m}{}^{ab}$ and $\o_\m{}^a$
\bea
\o_\m{}^{ab}&=& -2 e^{\n [a} \partial_{[\m} e_{\n]}{}^{b]} + e^{\n a} e^{\r b} e_{\m c} \partial_{[\n} e_{\r]}{}^c
- \t_\m e^{\n a} e^{\r b} \partial_{[\n} m_{\r]}  +  2e_{\mu}^{ \ [a}e_{\nu}^{\ b]}b^\nu\,,\nn\\
\o_\m{}^a  &=& \t^\n \partial_{[\m} e_{\n]}{}^a + e^{\n a} \t^\r e_{\m b} \partial_{[\n} e_{\r]}{}^b
+ e^{\n a} \partial_{[\m} m_{\n]} + \t_\m \t^\n e^{\r a} \partial_{[\n} m_{\r]} + e_{\m }^{\ a}\tau^\n b_\n \,
\label{OmegaOmega}
\eea

We can now define the connection for the scale invariant gravity in terms of the elements of the scale-extended Bargmann algebra by means of scale covariant metric compatibility conditions
\bea
0 &=& \partial_{\m} \t_{\n} - \G_{\m\n}^\r \t_{\r} - 2 b_\m \t_{\n} \,,\nn\\
0&=& \partial_{\m} e_\n{}^a - \G_{\m\n}^\r e_{\r}{}^{a} - \o_{\m}{}^{ab} e_{\r b} - \o_\m{}^a \t_{\n} - b_\m e_\n{}^a \,. 
\label{ScaleComp}
\eea
These conditions uniquely determine $\G$ as
\bea
{\G}_{\m\n}^\r = \t^\r \cD_\m \t_\n + \frac12 {h}^{\r\s} \Big( \cD_\n {h}_{\s\m} + \cD_\m  h_{\s\n} - \cD_\s  h_{\m\n} \Big) - h^{\r\s} \t_{(\m}F_{\n)\s}   \,,
\label{ScaleConn}
\eea
where the scale-covariant objects are as defined as 
\bea
 \cD_{\m} \t_{\n} = \partial_{\m} \t_{\n} - 2 b_{\m} \t_{\n}\,, \quad \cD_{\m} {h}_{\n\r} = \partial_{\m}  {h}_{\n\r} - 2 b_\m {h}_{\n\r} \,.
\eea
If desired, we can simply go to the ``hatted frame" by introducing a compensating scalar $\chi$, defining the vector $M_\m$ via
\bea
m_\m = M_\m +  \partial_\m \chi 
\eea
and finally using the map (\ref{OldNew}) in which case we obtain the connection given in (\ref{SConn}) for $z=2$ and $K_{\m\n}{}^\r  = 0$.
Finally, we give the corresponding scale invariant Riemann tensor in terms of $R_{\m\n}{}^{ab} (J), R_{\m\n}{}^a (G)$ and $R_{\m\n}(D)$ as
\bea
R^\r{}_{\m\n\s} (\G)= - e^\r{}_a \Big( \t_\m R_{\n\s}{}^a (G) + e_{\m b} R_{\n\s}{}^{ab}(J)   + e_{\m}{}^{a} R_{\n\s}(D) \Big) - 2 \t^\r \t_\m R_{\n\s}(D) \,.
\label{z2Riem}
\eea
As described in Section \ref{Section3}, we can use this scale invariant Riemann tensor to construct scale-invariant non-relativistic gravity actions or field equations by introducing a compensating scalar field $\phi$ and a rank-$(d+1)$ Milne-invariant tensor $g_{\m\n}$, see (\ref{gmn}).

\subsection{Schr\"odinger Algebra}
\paragraph{}
When the symmetries are extended to the Schr\"odinger extension of the Bargmann algebra, we include the non-relativistic analogue  the special conformal symmetry generator $K$
\bea
\{H\,, \quad P_a \,, \quad J_{ab} \,, \quad G_a \,, \quad N\,, \quad D \, \quad K \} \,
\eea
which introduces an additional gauge field $f_\m$. The commutation relations between the generators of the Schr\"odinger algebra is given by \cite{Bergshoeff:2015ija}
\begin{eqnarray}\label{Schrodingeralgebra}
&&[D,P_a] = -P_a\,,\hskip 1.3truecm [D,H] = -2H\,,\hskip 1.2truecm  [H,G_a] = P_a\,,\hskip 1.3truecm [P_a,G_b] = \delta_{ab}N\,,\nonumber\\[.2truecm]
&& [D,G_a] = G_a\,,\hskip 1.5truecm [D, K] = 2K\,,\hskip 1.5truecm [K, P_a] = -G_a\,,\hskip 1.0truecm [H,K] = D\,, \nonumber\\[.2truecm]
&&[J_{ab}, P_c] = 2\delta_{c[a}P_{b]}\,, \hskip 0.6truecm  [J_{ab}, G_c] = 2\delta_{c[a}G_{b]}\,, \hskip 0.6truecm  [J_{ab}, J_{cd}] = 4\delta_{[a[c}\,J_{b]d]}\,.
\end{eqnarray}
%Therefore, the full set of corresponding gauge fields in the case of Schr\"odinger symmetry is given by
%\bea
%h_\m{}^A &=& \{ \t_\m \,, \quad   e_\m{}^a \,, \quad \o_{\m}{}^{ab} \,, \quad \o_{\m}{}^a \,, \quad m_\m , \quad b_\m , \quad f_\m \} \,.
%\eea
Using the structure constants for the Schr\"odinger group \cite{Bergshoeff:2014uea}, we give the transformation rules for the gauge fields as
\bea
\d \t_\m &=& \partial_\m \xi + 2 \L_D \t_\m - 2 b_\m \xi \,,\nn\\
\d e_\m{}^a &=& \partial_\m \xi^a  - \o_\m{}^{ab} \x_b  - b_\m \x^a + \l^a{}_b e_\m{}^b + \l^a \t_\m - \o_\m{}^a \x + \L_D e_\m{}^a \,,\nn\\
\d \o_\m{}^{ab} &=& \partial_\m \l^{ab} + 2\l^{c[a} \o_\m{}^{b]}{}_c \,,\nn\\
\d \o_\m{}^a &=& \partial_\m \l^a + b_\m \l^a - \o_{\m}{}^{ab} \l_b - \o_\m{}^a \L_D - f_\m \x^a + e_\m{}^a \L_K  + \l^{a}{}_{b}\,  \o_{\m}{}^{b}  \,,\nn\\
\d b_\m &=& \partial_\m \L_D + \t_\m \L_K - \x f_\m \,,\nn\\
\d f_\m &=& \partial_\m \L_K + 2 \L_K b_\m - 2 f_\m \L_D \,,\nn\\
\d m_\m &=& \partial_\m \s + \l^a e_{\m a} + \o_\m{}^a \x_a \,.
\label{SchTrans}
\eea
while  the corresponding curvatures are \cite{Bergshoeff:2014uea}
\bea
R_{\m\n}(H) &=& 2 \partial_{[\m} \t_{\n]} - 4 b_{[\m} \t_{\n]} \,,\nn\\
R_{\m\n}{}^a (P) &=&  2 \partial_{[\m} e_{\n]}{}^a - 2 \o_{[\m}{}^{ab} e_{\n]b} - 2 \o_{[\m}{}^a \t_{\n]} - 2 b_{[\m} e_{\n]}{}^a \,,\nn\\
R_{\m\n}{}^{ab} (J) &=& 2 \partial_{[\m} \o_{\n]}{}^{ab} - 2 \o_{[\m}{}^{c[a} \o_{\n]}{}^{b]}{}_c \,,\nn\\
R_{\m\n}{}^a (G) &=& 2 \partial_{[\m} \o_{\n]}{}^a + 2 \o_{[\m}{}^{b} \o_{\n]}{}^a{}_b - 2 \o_{[\m}{}^{a} b_{\n]} - 2 f_{[\m}{} e_{\n]}{}^a \,,\nn\\
R_{\m\n} (D) &=& 2 \partial_{[\m} b_{\n]} - 2 f_{[\m} \t_{\n]} \,,\nn\\
R_{\m\n} (K) &=& 2 \partial_{[\m} f_{\n]} + 4 b_{[\m} f_{\n]} \,,\nn\\
R_{\m\n} (N) &=& 2 \partial_{[\m} m_{\n]} - 2\o_{[\m}{}^a e_{\n]a} \,.
\eea

In order to make contact to the conformal generalization of the Newton-Cartan geometry that we mentioned in the previous section, we want to solve $\o_\m{}^{ab}, \o_{\m}{}^a, e^\m{}_a \ , b_\m$ and $f_\m$ in terms of the other fields to leave $\t_{\m}, e_\m{}^a, m_\m$ and $\t^\m b_\m$  as the set of fields that characterizes the conformal extension of the Newton-Cartan geometry. This can be achieved in two distinct way.
\begin{itemize}
\item {In the first case, we impose the following set of constraints
\bea
&& R_{\m\n}(H) = 0 \,, \quad R_{\m\n}{}^a (P) = 0 \,, \quad R_{\m\n} (N) = 0 \,,\nn\\
&& R_{\m\n} (D) = 0 \,, \quad e^\m{}_a \t^\n R_{\m\n}{}^a (G) = 0 \,, \quad  R_{\m\n}{}^{ab} (J) = 0 \,.
\eea
These constraints imply further constraints due to Bianchi identities
\bea
R_{[abc]} (G) = 0 \,, \quad R_{0[ab]}(G) = 0 \,, \quad R_{ab}{}^c (G) = 0 \,, \quad R_{ab} (K) = 0  \,.
\eea
From these constraints, we find the solutions of $\o_\m{}^{ab}, \o_{\m}{}^a, e^\m{}_a \, b_\m$ and $f_\m$ as
\bea
&& \o_{\m}{}^{ab}  =0 \,, \quad \t^\m f_\m = \frac1{d}  R^\prime_{0a}{}^a (G)  \,,\quad e^\m{}_a f_\m = 2 e^\m{}_a \t^\n \partial_{[\m} b_{\n]} \,, \quad e^\m{}_a b_\m = e^\m{}_a \t^\n \partial_{[\m} \t_{\n]}  \,,\nn\\
&& \o_\m{}^a  = \t^\n \partial_{[\m} e_{\n]}{}^a + e^{\n a} \t^\r e_{\m b} \partial_{[\n} e_{\r]}{}^b
+ e^{\n a} \partial_{[\m} m_{\n]} + \t_\m \t^\n e^{\r a} \partial_{[\n} m_{\r]} + e_{\m }^{\ a}\tau^\n b_\n \,.
\eea
where 
\bea
R^\prime_{\m\n}{}^a (G) = 2 \partial_{[\m} \o_{\n]}{}^a + 2 \o_{[\m}{}^{b} \o_{\n]}{}^a{}_b - 2 \o_{[\m}{}^{a} b_{\n]} \label{f0}
\eea
}
\item {In the second case, we first introduce a scalar field $\chi$ and define the $\U(1)$ invariant vector $M_\m$ as in (\ref{MmChi}). Next, we impose the following set of constraints \cite{Bergshoeff:2014uea}
	\bea
&& R_{\m\n}(H) = 0 \,, \quad R_{\m\n}{}^a (P) = 0 \,, \quad R_{\m\n} (N) = 0 \,, \quad R_{\m\n} (D) = 0\nn\\
&& \tau^\mu e^\nu{}_{a}R_{\mu\nu}{}^a(G) + M^b [ 2\tau^\mu e^\nu{}_a R_{\mu\nu}{}^{ab}(J)+ M^c e^\mu{}_{c} \,  e^{\nu}{}_{a} R_{\mu\nu}{}^a{}_{b} (J) ]  = 0	 \,.
\label{Constraints2}
	\eea
These constraints imply further constraints due to Bianchi identities \cite{Bergshoeff:2014uea}
\bea
&& R_{[abc]} (G) = 0\,, \quad R_{0[ab]}(G) = 0 \,, \quad R_{[abc]}{}^d (J) = 0 \,, \nn\\
&&  2 R_{0[ab]}{}^c (J) - R_{ab}{}^c (G) = 0 \,, \quad R_{ab} (K) = 0 \,.
\eea
Using these constraints, we find that the composite expressions for $\o_{\m}{}^{ab}$ and $\o_{\m}{}^a$ are still given by (\ref{OmegaOmega}) while the solutions for $e^\m{}_a \, b_\m$ and  $f_\m$ read \cite{Bergshoeff:2014uea}
\bea
\t^\m f_\m &=& \frac1{d} \Big( \t^\m e^{\n}{}_a R^\prime_{\m\n}{}^a (G) + 2 \t^\m M^c R_{\m a}{}^a{}_c (J) + M^b M^c R_{ba}{}^a{}_c (J) \Big) \,,\nn\\
 e^\m{}_a f_\m &=& 2 e^\m{}_a \t^\n \partial_{[\m} b_{\n]}    \,, \quad e^\m{}_a b_\m = e^\m{}_a \t^\n \partial_{[\m} \t_{\n]} \,.
\eea
}
\end{itemize}
As shown in \cite{Afshar:2015aku}, the second case is the most convenient to construct Schr\"odinger invariant gravity models as the first case sets $\o_\m{}^{ab} = 0$ by imposing $R_{\m\n}{}^{ab} (J) = 0$. Thus, we will utilize the second set of constraints. With $e_\m{}^a, \t_\m, \t^\m b_\m$ and $M_\m$ being the independent fields of the Schr\"odinger invariant non-relativistic gravity, we observe that $\t^\m b_\m$ is the only independent field that transforms non-trivially under $K-$transformations as in the geometric construction (\ref{TransformB}). Furthermore, the scale-covariant compatibility conditions (\ref{ScaleComp}) remains unchanged in the presence of the special conformal symmetry, thereby the connection is again given by (\ref{ScaleConn}) which has a non-trivial $K-$transformation due to the appearance of $\t^\m b_\m$ terms in its definition as shown in (\ref{GammaB}). 
%
%%%%%%%%%%%%%%%%%%%%%%%%%%%%%%%%%%%%%%%%%%%%%%%%%%%%%%%%%%%%%%%%%%
%\section{Non-Relativistic Scale Invariant and Schr\"odinger Gravity}{\label{Sec3}}
%%%%%%%%%%%%%%%%%%%%%%%%%%%%%%%%%%%%%%%%%%%%%%%%%%%%%%%%%%%%%%%%%%

\subsection{Scale Invariant Non-Relativistic Gravity}{\label{Sec1}}
\paragraph{}
The purpose of this section is to classify the scale invariant scalar field theories that are relevant to the Ho\v{r}ava-Lifshitz gravity after gauge fixing of the scale symmetry. In order to do so, we introduce two scalar fields $\phi$ and $\chi$, which has the following transformations under scale and $\U(1)$ transformations
\bea
\d \phi = \o \L_D \phi \,, \qquad \d \chi = \s \,,
\eea
where $\o$ refers to the scaling dimension of the scalar field $\phi$. Note that we purposefully used $\phi$ and $\chi$ to refer to the scalars that we used in the geometric construction of the scale and Schr\"odinger invariant models, see (\ref{Phi}) and (\ref{Chi}). Furthermore, we will also use a complex scalar field $\Psi$ that is defined in terms of $\phi$ and $\chi$ as
\bea
\Psi = \phi  \, e^{{\rm i } \mass  \chi} \,,
\label{Psi}
\eea
which transforms homogeneously under dilatations and $\U (1)$ transformations
\bea
\d \Psi = \o \L_{D} \Psi + \rmi \mass \s \Psi \,. 
\eea
The models that we will introduce here fall in three classes. First two class are separated depending on the number of time derivatives acting on the scalar fields. The third class is necessary to introduce scale-invariant models that do not arise as a scalar field theory as they include group theoretical curvatures given in (\ref{ScaleCurv}).
\begin{itemize}
\item {\textbf{Potential Terms:} Lagrangians that are zeroth, second and fourth order in spatial derivatives $(n_s = 0, 2, 4)$, and no time derivative $(n_t = 0)$.}
\item {\textbf{Kintetic Terms:} Lagrangians that are first and second order in time derivative $(n_t = 1, 2)$.}
\item{\textbf{Curvature Terms:} Lagrangians that are constructed by using the group theoretical curvatures which we defined in (\ref{ScaleCurv}).}
\end{itemize}
Before we proceed to the construction of the scale invariant models, we would like to point out that any non-relativistic Lagrangian can be made scale invariant by using the above mentioned compensating fields. Here, the challenge is to find the class of ``any non-relativistic Lagrangians" which is not as trivial as the relativistic case due to Galilean transformations.

\subsubsection{Potential Terms}
\paragraph{}
In this section, we construct the Lagrangians that are zeroth, second and fourth order in spatial derivatives, while keeping the time derivative at zeroth order. 
\begin{itemize}
\item {$n_s = 0$: The only possibility for a scalar field theory that has no space or time derivatives is given by
\bea
S^{(0)} = \int dt\, d^d x\, e \, \L_0 \, \phi^2  
\eea 
where $\L_0$ is an arbitrary constant and $e = \det(\t_\m , e_\m{}^a)$ has the following scaling transformation
\bea
\d_D e = (d +2) \L_D \, e \,.
\eea 
Here, we also fixed the scaling dimension of the scalar field $\phi$ as $\o = - \frac{d+2}{2}$ for the scale invariance of the action.}
\item {$n_s = 2$: For this case, the only scale-invariant model is
\bea
S^{(1)} = \int dt\, d^d x\, e \, \cD_a \phi \, \cD^a \phi \,,
\eea
where $\cD_a \phi$ is the spatial part of the gauge-covariant derivative $\cD_\m \phi$
\bea
\cD_a \phi = e^\m{}_a \cD_\m \phi  \qquad \text{where} \qquad \cD_\m \phi = \partial_{\m} \phi - \o b_\m \phi \,.
\eea
In this case the scaling dimension of the scalar field is fixed to $\o = -\frac{d}{2}$. In principle, one can also have an action that includes $\phi \cD^a \cD_a \phi$ term. However, as noted in \cite{Afshar:2015aku}, it is related to $S^{(1)}$ up to a boundary term.}
\item {For $n_s = 4$, we have three distinct actions that contributes as potential terms
\bea
S^{(2)} &=& \int dt\, d^d x\, e \, \phi^{-2} \, (\cD_a \phi \, \cD^a \phi)^2 \,, \nn\\
S^{(3)} &=& \int dt\, d^d x\, e \, \phi^{-1} \, (\cD_a \phi \, \cD^a \phi) \triangle \phi  \,,\nn\\ 
S^{(4)} &=& \int dt\, d^d x\, e \, (\triangle \phi)^2  \,,
\eea
where 
\bea
\triangle \phi = \cD^a \cD_a \phi = e^{\m a} \Big[ (\partial_\m - (\o-1) b_\m ) \cD_a \phi - \o_{\m a}{}^b \, \cD_b \phi  \Big]\,.
\eea 
Note that in this case, the scaling dimension of the scalar field $\phi$ as $\o = - \frac{d-2}{2}$. In principle, one can also have two other $n_s = 4$ actions given by $\phi \triangle^2 \phi$ and $\cD_a \cD_b \phi \cD^a \cD^b \phi$. However,as noted in \cite{Afshar:2015aku}, these terms can be written in terms of a combination of $n_s = 4$ and curvature terms up to boundary terms. Thus, we will not include them in our list of $n_s = 4$ actions.}
\end{itemize}

In principle, we can also produce potential terms with the scalar field $\chi$. However, as shown in \cite{Afshar:2015aku}, such potential terms arises in the kinetic terms of a complex scalar field, which we will consider in the next section. Thus, these actions complete our list of potential terms.

%Note that we did not include potential terms for $\chi$ as will arise as a part of kinetic terms that we will introduce in the next subsection.

\subsubsection{Kinetic Terms}
\paragraph{}
In this section, we construct the actions that are first and second order in time derivative by utilizing the complex scalar field $\Psi$ that we introduced in (\ref{Psi}). 
\begin{itemize}
\item {At first order in time derivative, $n_t = 1$, we have
	\bea
	S^{(5)} &=& \int dt\, d^d x\, e \, \Psi^\star \Box \Psi  \,,
	\eea
	Here, we defined the scale-covariant d'Alambertian operator that is given by
	\bea
	\Box \Psi = \Big( \rmi  \cD_0 - \tfrac{1}{2\mass} \triangle  \Big) \Psi \,
	\eea
where 
\bea
 \cD_\m \Psi &=& \partial_\m \Psi - \o b_\m \Psi - \rmi \mass m_\m \Psi  \,. \nn\\
\triangle \Psi &=& e^{\mu a } \Big((\partial_\mu - (\omega -1)b_\mu  - \rmi \mass m_\mu )\cD_a \Psi  - \omega_{\mu a}{}^{c} \cD_c \Psi+ \rmi  \mass \omega_{\mu a}\Psi \Big) \,,
\eea
Note that the scale invariance fixes the scaling dimension of $\P$ as $\o = -\frac{d}{2}$ while $\U(1)$ charge of $\P$ remains arbitrary.}
\item {$n_t = 2$: For this case, we have the following set of invariant modes
\bea
S^{(6)} &=& \int dt\, d^d x\, e \, \Psi^\star \Box^2 \Psi  \,,\nn\\
S^{(7)} &=& \int dt\, d^d x\, e \, | \Box \Psi |^2\,,\nn\\
S^{(8)} &=& \int dt\, d^d x\, e \, \Big| \triangle \Psi - \frac{\cD^a \Psi \cD_a \Psi}{\Psi} \Big|^2  \,,\nn\\
S^{(9)} &=& \int dt\, d^d x\, e \, (\Psi^\star \Psi)^{-1} \Big( \rmi \Psi^\star \cD_0 \Psi - \rmi \Psi \cD_0 \Psi^\star + \frac{1}{\mass} \cD_a \Psi^\star \cD^a \Psi \Big)^2 \,, \nn\\
S^{(10)} &=& \int dt\, d^d x\, e \, \phi^{-2} \cD_a \phi \cD^a \phi \Big( \rmi \Psi^\star \cD_0 \Psi - \rmi \Psi \cD_0 \Psi^\star + \frac{1}{\mass} \cD_a \Psi^\star \cD^a \Psi \Big) \,,\nn\\
S^{(11)} &=& \int dt\, d^d x\, e \, \f^{-1} \triangle \f \Big( \rmi \Psi^\star \cD_0 \Psi - \rmi \Psi \cD_0 \Psi^\star + \frac{1}{\mass} \cD_a \Psi^\star \cD^a \Psi \Big) \,.
\eea 
where we have defined
\bea
\cD_0^2 \Psi &=& \tau^\mu\Big((\partial_\mu - (\omega -2)b_\mu  -\rmi  \mass m_\mu )\cD_0 \Psi  + \omega_\mu{}^a \cD_a \Psi \Big) \,,\nn\\
\cD_\m \triangle \Psi &=&  (\partial_\mu - (\omega -2)b_\mu  -\rmi \mass m_\mu )\triangle \Psi + 2\rmi \mass \omega_\mu{}^{a}\cD_a \Psi   \,,\nn\\
\triangle \cD_0 \Psi &=&e^{\mu}_{\ a}\Big((\partial_\mu - (\omega -3)b_\mu  -i\mass m_\mu )\cD_a \cD_0 \Psi - \omega_{\mu a}{}^{b}\cD_b \cD_0 \Psi\nn\\
&& \qquad + \omega_\mu{}^{b}\cD_a \cD_b\Psi  + i\mass \omega_{\mu a}\cD_0 \Psi \Big) \,,\nn\\
\triangle^2\Psi &=& e^{\m a}  \Big( (\partial_\mu - (\omega -3)b_\mu  -i\mass m_\mu )\cD_a\triangle \Psi - \omega_{\mu a}{}^{b} \cD_b \triangle \Psi \nn\\
&& \qquad + 2\rmi \mass \omega_\mu{}^{b}\cD_a \cD_b \Psi   + \rmi \mass \omega_{\mu a}\triangle \Psi \Big)\,,\nn\\ 
\Box^2 \Psi &=& \Big( - \cD_0^2 + \ft{1}{4\mass^2} \triangle^2  - \ft{\rmi}{2\mass} \cD_0 \triangle - \ft{\rmi}{2\mass} \triangle  \cD_0 \Big) \Psi  \,.
\eea}
\end{itemize}
Note that the action $S^{(8)}$ is actually a potential term for the complex scalar field $\P$, however we include it in the kinetic actions for future purposes. The order of derivatives matter in the Lagrangians due to the following non-vanishing commutation relations
\bea
 \left[\cD_0 , \triangle \right] \Psi  &=& - \o \cD^a R_{0a} (D) \P - (2\o - 1) R_{0a}(D) \cD^a \P + R_{0ab}{}^a (J) \cD^b \P + \rmi \mass R_{0a}{}^a (G) \P \,,\nn\\
 \left[\cD_0 , \cD_a \right] \Psi &=& - \o R_{0a}(D) \Psi \,.
\eea
Here, the scale invariance of the action fixes the scaling dimension of both the real scalar $\f$ and the complex scalar field $\Psi$ as $\o = - \tfrac{d-2}{2}$. 

\subsubsection{Curvature Terms}
\paragraph{}
In this section, we consider the curvature invariants, i.e., Lagrangians that are constructed by using the group theoretical curvatures. Such actions cannot be obtained by gauging a globally scale-invariant scalar field theory, thus deserved a special attention \cite{Afshar:2015aku}. In particular, the following two actions play an important r\'ole in comparing scale and Schr\"odinger invariant models
\bea
S^{(12)} &=& \int dt\, d^d x\, e \, \Big(\tau^\mu e^\nu{}_{a}R_{\mu\nu}{}^{a}(G) + M^b[2\tau^\mu e^\nu{}_{a} R_{\mu\nu}{}^{ab}(J)\nn\\
&& \qquad \qquad \quad - d  \tau^\mu e_\nu{}^{b} R_{\mu\nu}(D) + M^c e^\mu{}_{c}e^{\nu}{}_{a} R_{\mu\nu}{}^{a}{}_b (J)]\Big) (\P \P^\star)  \,,\nn\\
S^{(13)} &=& \int dt\, d^d x\, e \, \Big( -\frac{\rmi}{2\mass}\Big[  \tau^\mu e^\nu{}_a R_{\mu\nu}(D) \cD^a \P + \cD^a(\tau^\mu e^\nu{}_a R_{\mu\nu}(D) \P ) \Big]   \nn\\
&& \qquad \qquad \quad + M^a e^\nu{}_a \tau^\mu R_{\mu\nu}(D) \P \Big)  \P^\star
\eea
where the scaling dimension of the complex scalar $\P$ is given by $\o = - \ft{d-2}{2}$.  We can construct further independent invariants that cannot be obtained from a scalar field theory by using the spatial rotation curvature $R_{abcd} (J)$ \cite{Afshar:2015aku}
\bea
S^{(14)} &=& \int dt\, d^d x\, e \,  R(J) \phi^2 \nn\\
S^{(15)} &=& \int dt\, d^d x\, e \,  R(J)^2 \phi^2 \nn\\
S^{(16)} &=& \int dt\, d^d x\, e \,  \cD_a \phi \cD^a \phi \, R(J) \nn\\
S^{(17)} &=& \int dt\, d^d x\, e \, \phi \triangle \f \, R(J) \nn\\
S^{(18)} &=& \int dt\, d^d x\, e \, \cD_a \f \cD_b \f R^{ab} (J) \nn\\
S^{(29)} &=& \int dt\, d^d x\, e \, \f^2 R_{ab}(J) R^{ab}(J) \nn\\
S^{(20)} &=& \int dt\, d^d x\, e \,  \f^2 R_{abcd} (J) R^{abcd}(J) \,,\nn\\
S^{(21)} &=& \int dt\, d^d x\, e \,  \f^2 \Big( \rmi \Psi^\star \cD_0 \Psi - \rmi \Psi \cD_0 \Psi^\star + \frac{1}{\mass} \cD_a \Psi^\star \cD^a \Psi \Big) R(J) \,.
\label{RJSch}
\eea
where 
\bea
R(J) \equiv R_{ab}{}^{ab}(J)\,, \qquad \text{and} \qquad R_{ab} (J) \equiv R^{c}{}_{acb} (J) \,.
\eea
Finally, we give the scale invariant action for $R_{\m\n}(D)$, which is, as we will discuss below, a crucial action to distinguish scale invariance from Schr\"odinger invariance
\bea
S_D  &=& \int dt\, d^d x\, e \, \f^2 \, \hat\t^\m \hat{\t}^\r e^\n{}_a e^{\s a} R_{\m\n} (D) R_{\r\s}(D) \,.
\label{RD}
\eea
where the scaling dimension of the real scalar $\f$ is given by $\o = \frac{4-d}{2}$.

\subsection{Schr\"odinger Gravity}
\paragraph{}
We now have all the desired actions to proceed to the Schr\"odinger gravity models and compare them with the scale-invariant non-relativistic gravity theories. Since the difference between the scale and the Schr\"odinger symmetry is the existence of $\t^\m b_\m$ terms, it is worth mentioning such terms in the scale-invariant models that we constructed above.
\begin{itemize}
\item {\textbf{Potential Terms:} None of the potential terms constructed here includes a $\t^\m b_\m$ terms. Thus, they exhibit a Schr\"odinger invariance.}
\item {\textbf{Kinetic Terms:} $S^{(6)}, S^{(7)}$ and $S^{(8)}$  fail the Schr\"odinger invariance as they include explicit $\t^\m b_\m$ terms. Other kinetic actions exhibit a Schr\"odinger invariance.}
\item {\textbf{Curvature Terms:} Only $S^{(12)}$ and $S^{(13)}$ fails the Schr\"odinger invariance as they include        explicit $\t^\m b_\m$ terms. Other curvature actions exhibit a Schr\"odinger invariance.}
\end{itemize}

As the scale invariant potential terms are already Schr\"odinger invariant, we start our investigation with the kinetic terms. For an arbitrary scaling dimension of the complex scalar $\P$, the $b_0 \equiv \t^\m b_\m$ terms in the terms relevant to $S^{(6)}, S^{(7)}$ and $S^{(8)}$ are given by
\bea
\label{BoxBox}
\P^\star \Box^2 \P |_{\t^\m b_\m} &=&  - \rmi (2\omega -2 + d)b_0 ({\P}^\star \Box  \P) -\frac{\rmi}{2\mass}(\omega + \frac{d}{2}){\P}^\star \cD_a \P  (\tau^\mu e^\nu{}_a R_{\mu\nu}(D))\nn \\
&& -\frac{\rmi }{2\mass}(\omega + \frac{d}{2}){\P}^\star \cD^a(\tau^\mu e^\nu{}_a R_{\mu\nu}(D)\P) + (\omega + \frac{d}{2}) (\P^\star \P) \partial_0 b_0 \nn\\
&& + (\omega + \frac{d}{2})^2 (\P^\star \P) b_0^2 \,,\\
\Box \P |_{\t^\m b_\m} &=&  - \rmi (\o + \ft{d}{2}) b_0 \P \,,\\
\triangle \P|_{\t^\m b_\m} &=& \rmi \mass d b_0 \P \,.
\eea
From these expressions, we first observe that given the scaling dimension of $\P$ is $\o = - \tfrac{d-2}{2}$, the following combination does not have an explicit $b_0$ term and is invariant under the full scale-extended Bargmann group, thus exhibits a Schr\"odinger invariance
\bea
S  &=& \int dt\, d^d x\, e \, \Big| \Box \P + \frac{1}{\mass d} \Big(\triangle \Psi - \frac{\cD^a \Psi \cD_a \Psi}{\Psi} \Big) \Big|^2  \,.
\eea
This action is precisely the one of the Schr\"odinger invariant actions found in \cite{Afshar:2015aku}. Thus, in order to distinguish between the scale and Schr\"odinger invariance, we can introduce a parameter $\a$ that measures the deviation from Schr\"odinger symmetry while preserving the scale invariance
\bea
S^{(22)}  &=& \int dt\, d^d x\, e \, \Big| \Box \P + \frac{\a}{\mass d} \Big(\triangle \Psi - \frac{\cD^a \Psi \cD_a \Psi}{\Psi} \Big) \Big|^2  \,.
\eea
Here, when $\a = 1$ we have a full Schr\"odinger invariance, otherwise the model exhibits only scale invariance. 

For the $b_0$ terms in $\P^\star \Box^2 \P$, it is not possible to find a vanishing combination by use of only kinetic terms and choosing a weight, thus we turn our attention to the curvature actions $S^{(12)}$ and $S^{(13)}$. Given the scaling dimension of the complex scalar field $\P$ is $\o = - \tfrac{d-2}{2}$ we first note that the first expression in (\ref{BoxBox}) drops out, and the second and the third expressions can be compensated by $S^{(13)}$. Furthermore, the $b_0$ terms in $S^{(12)}$ is given by
\bea
S^{(12)} |_{\t^\m b_\m} &=& \int dt\, d^d x\, e \,  (\P^\star \P)  \Big( d (\partial_0 b_0 + b_0^2) - d M^a \t^\m e^\n{}_a R_{\m\n} (D) \Big) \,.
\eea
which has the correct structure to cancel out the last two $b_0$ structure in (\ref{BoxBox}). Thus, we find that the following scale-invariant combination also exhibits a Schr\"odinger invariance
\bea
S = S^{(6)} - \frac{1}{d} ( S^{(12)} + d S^{(13)} ) \,.
\eea
Once again, we make the distinction between the scale and Schr\"odinger invariance explicit by introducing a free parameter $\a$ such that
\bea
S^{(23)} = S^{(6)} - \frac{\a}{d} ( S^{(12)} + d S^{(13)} ) \,.
\label{BoxSqr}
\eea
If $\a \neq 1$, the the action (\ref{BoxSqr}) preserves the scale symmetry but no longer invariant under the special conformal transformations. When $\a=1$, the model reduces to the following Schr\"odinger invariant action \cite{Afshar:2015aku}
\bea
S|_{\a=1} = \int dt\, d^d x\, e \,  \P^\star \Box_{\rm Sch}^2 \Phi \,,
\label{BoxBoxSch}
\eea
where the $\Box^2_{\rm Sch}$ is the square of the Schr\"odinger invariant d'Alambertian operator
\bea
\Box^2_{\rm Sch} \P &=& \Big( - D_0^2 + \ft{1}{4\mass^2} \triangle_{\rm Sch}^2  - \ft{\rmi}{2\mass} D_0 \triangle_{\rm Sch} - \ft{\rmi}{2\mass} \triangle_{\rm Sch}  D_0 \Big) \Psi \,,
\eea
where the Schr\"odinger invariant derivative operators read \cite{Afshar:2015aku}
\bea
D_\m \P &=& ( \partial_{\m}  - \o b_\m  - \rmi \mass m_\m ) \P  \,,\nn\\
D_\m D_0 \P  &=& \Big( (\partial_\mu - (\omega -2)b_\mu  -i\mass m_\mu )D_0   + \omega_\mu{}^{a}D_a    + \o f_\m \Big) \P \nn \\
D_0^2 \P &=& \tau^\mu\Big((\partial_\mu - (\omega -2)b_\mu  -i \mass m_\mu )D_0  + \omega_\mu{}^{a}D_a  + \o f_\m \Big)  \P  \nn \\
D_aD_b \P  &=& e^{\mu}_{\ a}\Big((\partial_\mu - (\omega -1)b_\mu  -\rmi \mass m_\mu )D_b  - \omega_{\mu b}{}^{c}D_c + \rmi \mass \omega_\mu{}^{b}  \Big) \P  \nn \\
\triangle_{\rm Sch} \P &=& e^{\mu}_{\ a}\Big((\partial_\mu - (\omega -1)b_\mu  -i \mass m_\mu )D^a   - \omega_{\mu}{}^{ac} D_c + i \mass \omega_\mu{}^{a}  \Big) \P \nn \\
%D_aD_aZ &=& e^{\mu}_{\ a}\bigg((\partial_\mu - (\omega -1)b_\mu  -icA_\mu )D_aZ  - \omega_{\mu a}^{\ \ c}D_cZ+ ic\omega_{\mu a}Z \bigg) \nn \\
D_\m \triangle_{\rm Sch}  \P &=& \Big((\partial_\mu - (\omega -2)b_\mu  -i\mass m_\mu )\triangle_{\rm Sch}   + 2\rmi \mass \omega_\mu{}^{a}D_a - \rmi \mass d f_\m \Big) \P  \nn \\
\triangle_{\rm Sch}  D_0 \P &=& e^{\mu}_{\ a}\Big((\partial_\mu - (\omega -3)b_\mu  -\rmi \mass m_\mu )D_a D_0  - \omega_{\mu a}{}^{b}D_b D_0 + \omega_\mu{}^{b}D_aD_b \nn\\
&& \qquad + \rmi \mass \omega_{\mu a}D_0 + (\o-1) f_\m D_a \Big) \P  \nn\\
\triangle_{\rm Sch}^2 \P &=& e^{\m a} \Big[ (\partial_\mu - (\omega -3)b_\mu  - \rmi \mass m_\mu )D_a \triangle_{\rm Sch}  - \omega_{\mu a}{}^{b}D_b\triangle_{\rm Sch} \nn\\
&& \qquad + 2\rmi \mass \omega_\mu{}^{b}D_a D_b + \rmi \mass \omega_{\mu a}\triangle_{\rm Sch} - \rmi \mass (d+2) f_\m D_a  \Big] \P \,.
\label{SchCov}
\eea
Note that we utilized the gauge fields of the Schr\"odinger algebra (\ref{SchTrans}) to define the Schr\"odinger covariant derivatives (\ref{SchCov}). Here, the major difference between the scale and Schr\"odinger covariant objects is the existence of the composite $f_\m$ field. In the case of Schr\"odinger invariance, the composite $f_\m$ comes with a fixed coefficient to cancel out the $b_0$ terms to preserve the special conformal symmetry. In the case of scale invariance, the combination of $S^{(12)}$ and $S^{(13)}$ as given in (\ref{BoxSqr}) plays the role of $f_\m$. Hence, when $\a = 1$ that combination completes the scale covariant d'Alambertian-squared action $S^{(6)}$ to the Schr\"odinger covariant d'Alambertian-squared action (\ref{BoxBoxSch}), otherwise the model only exhibits scale invariance but not special conformal invariance.

We finish this section with a comment on the necessity of the action (\ref{RD}). If this action is not present, then $b_0$ becomes an auxiliary field and can simply be eliminated by its field equation. Thus, due to this elimination, any $z=2$ scale invariant model becomes Schr\"odinger invariant. Hence, when a model that aims to distinguish local scale invariant models from Schr\"odinger gravity, one must add this action with a coefficient $(\a-1)$ such that when $\a \neq 1$, then $b_0$ cannot be eliminated by its field equation, and when $\a=1$, the action  (\ref{RD}) drops out from the model in hand along with all other $b_0$ terms, giving rise to a Schr\"odinger invariant gravity.

%%%%%%%%%%%%%%%%%%%%%%%%%%%%%%%%%%%%%%%%%%%%%%%%%%%%%%%%%%%%%%%%%
\section{$z\neq 2$ Scale Invariant Ho\v{r}ava-Lifshitz Gravity}{\label{Sec4}}
\paragraph{}
In the previous section, we constructed all the $z=2$ scale invariant gravity models that are relevant to the Ho\v{r}ava-Lifshitz gravity and put an explicit distinction between the scale and Schr\"odinger invariant extension of the Ho\v{r}ava-Lifshitz gravity. In this section, our purpose is to develop a $z\neq2$ scale invariant tensor calculus  and construct the potential, kinetic and curvature terms that are relevant to the $z\neq2$ scale extension of the Ho\v{r}ava-Lifshitz gravity. Finally, following \cite{Hartong:2015zia}, we identify the $z\neq2$ scale extended Ho\v{r}ava-Lifshitz gravity. 

As mentioned in Section \ref{Section3}, when $z\neq2$, the scale symmetry can no longer be extended to the Schr\"odinger symmetry by including a non-relativistic special conformal transformation. Thus, the $z\neq2$ scale extended Bargmann algebra has the same generators and the gauge fields as in the $z=2$ scale extended case.The commutations relations between the generators of $z\neq2$ scale extended Bargmann algebra are given by \cite{Bergshoeff:2014uea}
\bea
&&[D,P_a] = -P_a\,,\qquad   [D,H] = -zH\,,\qquad  [H,G_a] = P_a\,,\nonumber\\[.2truecm]
&& [P_a,G_b] = \delta_{ab}N\,, \qquad [D,G_a] = (z-1) G_a\,, \qquad  [D,N] = (z-2) N \,, \nn\\
&& [J_{ab}, P_c] = 2\delta_{c[a}P_{b]}\,, \qquad [J_{ab}, G_c] = 2\delta_{c[a}G_{b]}\,, \qquad  [J_{ab}, J_{cd}] = 4\delta_{[a[c}\,J_{b]d]}\,.
\eea
The transformation rules are given by \cite{Bergshoeff:2014uea}
\bea
\d \t_\m &=& \partial_\m \xi - z \xi b_\mu +  z \L_D \t_\m\,,\nn\\
\d e_\m{}^a &=& \partial_\m \xi^a  - \o_\m{}^{ab} \x_b  - b_\m \x^a + \l^a{}_b e_\m{}^b + \l^a \t_\m - \o_\m{}^a \x + \L_D e_\m{}^a \,,\nn\\
\d \o_\m{}^{ab} &=& \partial_\m \l^{ab} + 2\l^{c[a} \o_\m{}^{b]}{}_c \,,\nn\\
\d \o_\m{}^a &=& \partial_\m \l^a - \o_{\m}{}^{ab} \l_b + \l^{a}{}_{b}\,  \o_{\m}{}^{b} + (z-1)\l^a b_\m - (z-1)\L_D \o_\m^{\ a} \,,\nn\\
\d m_\m &=& \partial_\m \s - \x^a \o_{\m a} + \l^a e_{\m a}  + (z-2) \s b_\m - (z-2) \L_D m_\m \,, \nn \\
\d b_\m &=& \partial_\m \L_D \,,
\label{zneqSITrans}
\eea
and the corresponding curvatures are given by \cite{Bergshoeff:2014uea}
\bea
R_{\m\n}(H) &=& 2 \partial_{[\m} \t_{\n]} - 2z b_{[\m} \t_{\n]} \,,\nn\\
R_{\m\n}{}^a (P) &=&  2 \partial_{[\m} e_{\n]}{}^a - 2 \o_{[\m}{}^{ab} e_{\n]b} - 2 \o_{[\m}{}^a \t_{\n]} - 2 b_{[\m} e_{\n]}{}^a \,,\nn\\
R_{\m\n}{}^{ab} (J) &=& 2 \partial_{[\m} \o_{\n]}{}^{ab} - 2 \o_{[\m}{}^{c[a} \o_{\n]}{}^{b]}{}_c \,,\nn\\
R_{\m\n}{}^a (G) &=& 2 \partial_{[\m} \o_{\n]}{}^a + 2 \o_{[\m}{}^{b} \o_{\n]}{}^a{}_b - 2 (z-1) \o_{[\m}{}^{a} b_{\n]} \,,\nn\\
R_{\m\n} (D) &=& 2 \partial_{[\m} b_{\n]}\,,\nn\\
R_{\m\n} (N) &=& 2 \partial_{[\m} m_{\n]} - 2\o_{[\m}{}^a e_{\n]a} + 2(z-2)b_{[\mu}m_{\nu]} \,.
\label{zneq2ScaleCurv}
\eea
Note that when the dynamical critical exponent $z$ is left arbitrary, the $D$ transformation rules for the inverse vielbein and the inverse temporal vielbein are given by
\bea
\d \t^\m =  - z \L_D \t^\m  \,, \qquad  \d e^\m{}_a =    - \L_D e^\m{}_a \,.
\eea
We are now at a position to make contact to the $z\neq2$ scale-invariant generalization of the Newton-Cartan geometry that we established in Section \ref{Section3}. As before, this is achieved imposing a set of curvature constraints. In the case $z\neq2$ scale symmetry, we have the following set of constraints \footnote{Our set of constraints is a bit different than that of \cite{Bergshoeff:2014uea}. In principle, we don't need to impose $R_{\m\n}(D) = 0$, but $R_{ab}(D) = 0$ is sufficient, which is the case studied in \cite{Bergshoeff:2014uea}. However, in that case the Bianchi identity for $R_{\m\n}(N)$ implies a gauge-dependent constraint: $e_{[\m}{}^a R_{\n\r]a}(G) = (z-2) m_{[\m} R_{\n\r]}(D)$, as long as $z\neq2$. Here, we avoid this gauge dependence by further imposing $R_{0a}(D) = 0$ which, together with $R_{ab}(D) = 0$ implies that $R_{\m\n}(D) = 0$.}
\bea
R_{\m\n}(H) = 0 \,, \qquad  R_{\m\n}{}^a (P) = 0 \,, \qquad R_{\m\n}(N) = 0 \,,  \qquad
R_{\m\n} (D) = 0 \,,
\label{CC3}
\eea
which results to the following further constraints by Bianchi identities
\bea
e_{[\m}{}^b R_{\n\r]}{}^{a}{}_{b} (J)  + \t_{[\m}R_{\n\r]}{}^a (G) = 0  \,,\qquad e_{[\m}{}^a R_{\n\r]a} (G) = 0 \,. \label{B3}
\eea
The first constraint implies that the twistless condition is satisfied, thereby the torsion vanishes 
\bea
R_{\m\n} (H)  = 0 \qquad \Rightarrow \qquad \partial_{[\mu} \t_{\n]}  = z b_{[\m} \t_{\n]} \,,
\label{RHSolve}
\eea
and determines the spatial part of $b_\m$ as
\bea
e^\m{}_a b_\m = \frac{2}{z}e^\mu{}_{a}\tau^\nu \partial_{[\mu}\tau_{\nu]} \,.
\eea
Furthermore, the last two constraints in (\ref{CC3}) gives rise to the solution of $\o_{\m}{}^{ab}$ and $\o_\m{}^a$
\bea
\o_\m{}^{ab}&=& -2e ^{\nu [a}\partial_{[\mu}e_{\nu]}{}^{b]}  + e^{\nu [a}e^{\ b] \rho}\partial_\nu e_\rho{}^{c} e_{\mu c} + 2e_{\mu}{}^{[a}e_{\nu}{}^{b]}b^\nu \nn\\
&&  -e^{\nu a}e^{\rho b}\tau_\mu( \partial_{[\nu}m_{\rho]} + (z-2)b_{[\n} m_{\r]} )\,,\nn\\
\o_\m{}^a  &=& \t^\n \partial_{[\m} e_{\n]}{}^a + e^{\n a} \t^\r e_{\m b} \partial_{[\n} e_{\r]}{}^b +e_{\m}{}^a \tau^\nu b_\nu + e^{\n a} (\partial_{[\m}m_{\n]} + (z-2)b_{[\m}m_{\n]}) \nn \\
&&+ \tau_\m \tau^\rho e^{\nu a}(\partial_{[\rho}m_{\nu]}+ (z-2)b_{[\rho}m_{\n]} ) \,.
\label{OmegaOmega2}
\eea
 To make contact with geometry, we turn to the $z \neq 2$ scale covariant metric compatibility conditions
 \bea
 0 &=& \partial_{\m} \t_{\n} - \G_{\m\n}^\r \t_{\r} - z b_\m \t_{\n} \,,\nn\\
 0&=& \partial_{\m} e_\n{}^a - \G_{\m\n}^\r e_{\r}{}^{a} - \o_{\m}{}^{ab} e_{\r b} - \o_\m{}^a \t_{\n} - b_\m e_\n{}^a \,. 
 \label{zneq2ScaleComp}
 \eea
 These conditions uniquely determine $\G$ as a symmetric connection
 \bea
 {\G}_{\m\n}^\r = \t^\r \cD_\m \t_\n + \frac12 {h}^{\r\s} \Big( \cD_\n {h}_{\s\m} + \cD_\m  h_{\s\n} - \cD_\s  h_{\m\n} \Big) - h^{\r\s} \t_{(\m}F_{\n)\s}  \,,
 \label{zneq2ScaleConn}
 \eea
where the $z\neq2$ scale-covariant objects are as defined as 
\bea
\cD_{\m} \t_{\n} = \partial_{\m} \t_{\n} - z b_{\m} \t_{\n}\,, \quad \cD_{\m} {h}_{\n\r} = \partial_{\m}  {h}_{\n\r} - 2 b_\m {h}_{\n\r} \,.
\eea
Finally, we give the corresponding $z\neq2$ scale invariant Riemann tensor in terms of $R_{\m\n}{}^{ab} (J)$ and $ R_{\m\n}{}^a (G)$ as
\bea
R^\r{}_{\m\n\s} (\G)= - e^\r{}_a \Big( \t_\m R_{\n\s}{}^a (G) + e_{\m b} R_{\n\s}{}^{ab}(J)   \Big) \,.
\label{zneq2Riem}
\eea
Once again, as described in Section \ref{Section3}, we can use this Riemann tensor to construct $z\neq 2$ scale-invariant non-relativistic gravity actions or field equations by introducing a compensating scalar field $\phi$ and a rank-$(d+1)$ Milne-invariant tensor $g_{\m\n}$.
\subsection{$z\neq 2$ Scale Invariant Tensor Calculus}
\paragraph{}
Our aim is to develop a tensor calculus to construct the $z\neq2$ generalization of the non-relativistic scale invariant gravity. As before, we are only interested in the set of models that are relevant to the Ho\v{r}ava-Lifshitz theory, thus we limit ourselves to a certain class of potential, kinetic and curvature terms. In principle, the construction procedure might seem like a straightforward generalization of what was done for the $z=2$ case. Furthermore, it seems natural to expect that the $z=2$ limit of the $z\neq2$ construction must recover the models that we give in Section \ref{Sec1}. Thus, before we start with the construction procedure, it is useful to enumerate the subtleties and technical differences of $z\neq2$ models.
\begin{enumerate}
\item {First of all, when $z\neq2$, the curvature of the dilatation gauge field, $R_{\m\n}(D)$ is set to zero due to Bianchi identity for $R_{\m\n}(N)$. Thus, $b_\m$ can be given  as $b_\m = \partial_\m \vf$ where $\vf$ is a scalar field that transforms as shift under dilatations $\d \vf = \L_D$ \footnote{If $R_{ab}(D) = 0$ is chosen as the constraint as in \cite{Bergshoeff:2014uea} instead of $R_{\m\n}(D) = 0$, then $b_\m$ can no longer be set to $b_\m = \partial_\m \vf$. In that case one can introduce a special conformal symmetry to the $z\neq 2$ scale extended Bargmann algebra by only taking the internal part of the algebra into account, which would lead one to a construction procedure similar to the construction of $z=2$ models given in Section \ref{Sec3}.}. This is certainly not the case for $z=2$ scale invariance, see (\ref{B2}).  }
\item {As $R_{\m\n}(D) = 0$ for $z\neq2$ models, the Riemann tensor for $z\neq2$ differs from the $z=2$ Riemann tensor by $R_{\m\n}(D)$ terms, see (\ref{z2Riem}) and (\ref{zneq2Riem}). Thus, there is no smooth $z=2$ limit of $z\neq2$ gravity theories that are constructed by use of the $z\neq2$ Riemann tensor. }
\item {$z\neq2$ scale-extended Bargmann algebra does not allow the existence of a scalar field with a homogeneous dilatation and $\U(1)$ transformation due to the following commutation relation
	\bea
	\left[D,N\right] = (z-2) N \,.
	\eea
Thus, as opposed to the $z=2$ case, we cannot introduce a complex scalar field as given in (\ref{Psi}). For $z\neq2$ setting we are only allowed to work with two type of scalar fields with the following transformation rules
\bea
\d \f = \o \L_D \f \,, \qquad \d \chi = \s + (2-z)\L_D \chi \,.
\eea }
\item {The scalar field $\chi$ has a non-vanishing $\U(1)$ transformation. This implies that we cannot form potential terms with $\chi$ and it can only appear in an action by its covariant derivative, which reads
\bea
\cD_{\m} \chi = \partial_\m \chi - (2-z) b_\m \chi - m_\m\,.
\eea 
This is the very definition of the $\U(1)$ invariant vector field $M_\m$ up to an overall sign difference, see (\ref{Mmubmu}). Thus, for the $z\neq 2$ setting, the main elements of the scale invariant tensor calculus are the scalar field $\f$ and the $\U(1)$ invariant vector field $M_\m$.}
\end{enumerate}
With these points in mind, we now proceed to the construction of the relevant potential, kinetic and curvature actions of $z\neq2$ scale invariant gravity.

\subsubsection{Potential Terms}
\paragraph{}
The transformation rules for $\f$ does not change in the  $z\neq2$ setting. Thus, the potential terms for $\f$ in $z\neq2$ scale-invariance is the same as the $z=2$ theory. On the other hand, unlike the $z=2$ case, we cannot define a complex scalar $\P$ field to include the potential terms of $\chi$ into the kinetic terms of the $\P$. Thus, here we give the potential terms of $\chi$ in zeroth, second and fourth order spatial derivatives.
\begin{itemize}
\item {$n_s = 0$: $\chi$ has a non-vanishing $\U(1)$ transformation, thus cannot form an action with no derivatives.}
\item {$n_s = 2$: For the construction of potential terms with two spatial derivatives, we turn to the spatial part of the covariant derivative of $M_a$
\bea
\cD_\m M_a = \partial_{\m} M_a + (z-1) b_\m M_a - \o_{\m}{}^{ab} M_b - \o_{\m a}  \,.
\eea
which is invariant under Galilean transformations due to the fact that $M_a$ transforms as shift under Galilean transformations
\bea
\d_G M_a = \l_a \,.
\eea
As the inverse spatial vielbein $e^\m{}_a$ is also Galilean invariant, we can form a spatial covariant derivative of $M_a$ that is invariant under Galilean transformations
\bea
\d \cD_a M_b = - z \L_D \cD_a M_b  \,.
\eea
Thus, the only possible $n_s = 2$ potential terms for $\chi$ are
\bea
S_{z\neq2}^{(1)} &=& \int dt\, d^d x\, e \, \f \, \cD_a M^a \,.
\eea
Here, the scaling dimension of $\f$ is given by $\o = -d$. Furthermore $e = \det(\t_\m , e_\m{}^a)$ has the following scaling transformation
\bea
\d_D e = (d +z) \L_D \, e \,.
\eea }
\item {$n_s = 4$: The construction of potential terms with four spatial derivatives can be divided into following three subclasses
\begin{enumerate}[a.]
\item {We first consider the models such that the spatial derivatives only act on $\chi$ terms ($n_{s,\chi} = 4, n_{s,\f} = 0$).
	\bea
	S_{z\neq2}^{(2)} &=& \int dt\, d^d x\, e \, \f \, (\cD_a M^a)^2 \,,\nn\\
	S_{z\neq2}^{(3)} &=& \int dt\, d^d x\, e \, \f \, (\cD_a M_b)^2  \,,\nn\\
	S_{z\neq2}^{(4)} &=& \int dt\, d^d x\, e \, \f \, \triangle \cD_a M^a  \,,
	\eea
	Here, for $S_{z\neq2}^{(2)}$ and $S_{z\neq2}^{(3)}$, the scaling dimension of $\f$ is given by $\o = z -d$ while for the $S_{z\neq2}^{(4)}$, the scaling dimension of $\f$ is $\o = 2-d$. Note that it is also possible to consider the models of kind $ \f \cD_a \triangle M^a$ or $\f \cD_a \cD_b \cD^a M^b$, however such actions are related to $S_{z\neq2}^{(4)}$ up to curvature invariants 
	\bea
	\cD_a  \triangle M^a &=&\triangle \cD_a M^a + \left[\cD_a,  \triangle  \right]M^a \nn\\
	&=&  \triangle \cD_a M^a  - \cD^b \Big( R_{bc}(J) M^c + R_{ab}{}^a (G) \Big) \,.
	\label{CurvDep}
	\eea
	We will construct that curvature invariant in the next section.}
\item {
	Next, we consider $n_{s,\chi} = 3, n_{s,\f} = 1$ models. In this case, the candidate models are given as
	\bea
	\cD_a \f \, \triangle M^a\,,\quad \cD^b \phi \cD_a \cD_b M^a\,, \quad \cD^b \phi \cD_b \cD_a M^a \,,
	\eea
	however, all these models are equivalent to $S_{z\neq2}^{(4)}$ up to boundary terms and the curvature term given in (\ref{CurvDep}).
%	we have the following possible potential term
%	\bea
%	S_{z\neq2}^{(5)} &=& \int dt\, d^d x\, e \, \cD_a \f \, \triangle M^a \,.
%	\eea
%	Here, the scaling dimension of $\f$ is given by $\o = 2 -d$. In principle, we could also have actions of type $\cD^b \phi \cD_a \cD_b M^a$ or $\cD^b \phi \cD_b \cD_a M^a$, however these actions are equivalent to $S_{z\neq2}^{(5)}$ up to a boundary term and the curvature term given in (\ref{CurvDep}).
}
\item {
	Finally, we consider the models with $n_{s,\chi} = 2, n_{s,\f} = 2$. In this case, the candidate models are given as
	\bea
	\cD_a M^a  \triangle \f \,, \quad \f^{-1} \cD_a M^a  \cD_b \f \cD^b \f \,, \quad \cD_a M_b \cD^a \cD^b \f \,, \quad \f^{-1}  \cD_a M_b \cD^a \f \cD^b \f \,.
	\eea
	However, all these models are equivalent to the previously constructed ones up to boundary terms. Thus, there is no independent $n_{s,\chi} = 2, n_{s,\f} = 2$ potential terms. }
\end{enumerate}
Note that as $M_a$ is not Galilean invariant, there is no $n_{s,\chi} = 1, n_{s,\f} = 3$ class of potential terms with. Furthermore, as mentioned before, $n_{s,\chi} = 0, n_{s,\f} = 4$ potential terms are the same as $z=2$ theory.} 
\end{itemize} 

\subsubsection{Kinetic Terms}
\paragraph{}
In this section, we construct the actions that are first and second order in time derivative. In order to do so, we first construct Galilean invariant quantities that include time derivatives on $\f$ or $\chi$. For the real scalar field $\f$, when no spatial derivative act on it, the only possible Galilean invariant quantity is
\bea
\cD_0 \f + M^a \cD_a \f \,.
\label{D0}
\eea
If we allow a single spatial derivative to act on $\cD_0 \f$ we also have a single Galilean invariant quantity
\bea
\cD_a \cD_0 \f + M^b \cD_a \cD_b \f \,.
\label{DD0f}
\eea
Note that in principle we could also have $\cD_0 \cD_a \f$, but it is the same $\cD_a \cD_0 \f$ since the commutator of $\cD_0$ and $\cD_a$ on the real scalar field vanishes. Next, we allow two spatial derivative to act on $\cD_0 \f$. In this case, there are two possible independent Galilean invariant quantities
\bea
\triangle \cD_0 \f  + M^a \triangle \, \cD_a \f \,, \qquad \cD_a \cD_b \cD_0 \f + M^c  \cD_a \cD_b \cD_c \f \,.
\eea
We could also have $\cD_0 \triangle \f$ and $\cD_a \cD_0 \cD^a \f$ but they are related to $\triangle \cD_0 \f$ by the curvature term $R_{0a}(J) \cD^a \f$. Another two possible actions at this level, $\cD_0 \cD_a \cD_b \f$ and $\cD_b \cD_0 \cD_a \f$, are related to $\cD_a \cD_b \cD_0 \f$ by the curvature term $R_{0ab}{}^{c}(J) \cD_c \f$.

When it comes to $\chi$ term, we work with the temporal component of the vector field $M_\m$. First, we don't allow a spatial derivative to act on $M_0$. In this case the only possible Galilean invariant quantity is given by
\bea
M_0 + \ft12 M_a M^a \,.
\eea
In the next step, we only allow a single spatial derivative to act on $M_0$, in which case the only Galilean invariant quantity is
\bea
\cD_a M_0 + M^b \cD_a M_b \,, 
\eea
Here, we could also have $\cD_0 M_a$ but it is equivalent to $\cD_a M_0$ as the commutator of $\cD_0$ and $\cD_a$ vanishes on  $\chi$. Finally, we allow two spatial derivative to act on $M_0$, in which case we have  two possible independent Galilean invariant quantities
\bea
\triangle M_0 + M^c \triangle M_c \,, \qquad \cD_b \cD_a M_0 + M^c \cD_b \cD_a M_c \,.
\eea
Note that we could also have $\cD_0 \cD_a M^a$ and $\cD_a \cD_0 M^a$ but they are related to $\triangle M_0$ up to the curvature term $R_{0a}{}^a (G) + R_{0a}{}^{ab}(J) M_b$. Furthermore, $\cD_0 \cD_a M_b$ and $\cD_b \cD_0 M_a$ are related to $\cD_a \cD_b M_0$ by the curvature term $R_{0a}{}^b (G) + R_{0a}{}^{bc}(J) M_c$.

When we have time derivatives at second order acting on $\f$ and $\chi$, we don't allow any spatial derivatives to act on such terms. In this case, the Galilean invariant quantities are
\bea
\cD_0^2 \f + 2 M^a \cD_0 \cD_a \f + M^a M^b \cD_a \cD_b \f   \,,\qquad \cD_0 M_0 + 2 M^a \cD_0 M_a + M^a M^b \cD_a M_b \,.
\eea
With these results in hand, we have the following classification of Galilean invariant actions.
\begin{itemize}
\item{$n_t = 1:$ When we have a single time derivative acting on $\f$ or $\chi$, we first consider the models with no spatial derivatives
\bea
S^{(5)}_{z\neq2} &=& \int dt\, d^d x\, e  \, \f (\cD_0 \f + M^a \cD_a \f) \,,\nn\\
S^{(6)}_{z\neq2} &=& \int dt\, d^d x\, e  \, \f^2 (M_0 + \ft12 M_a M^a) \,.
\eea
Here, for $S^{(5)}_{z\neq2}$ we have $\o = \ft{-d}{2}$, while for $S^{(6)}_{z\neq2}$ we have $\o = \ft{z-d-2}{2}$. The remaining $z\neq2$ scale-invariant actions, which consists one temporal and two spatial derivatives, can be classified with respect to the scaling dimension of the scalar field $\f$ as follows
\begin{enumerate}[a.]
	\item {For the following two models, the scaling dimension of the scalar field $\f$ is $\o = \ft{2-d}{2}$
	\bea
	S^{(7)}_{z\neq2} &=& \int dt\, d^d x\, e  \, \triangle \f (\cD_0 \f + M^a \cD_a \f) \,,\nn\\
	S^{(8)}_{z\neq2} &=& \int dt\, d^d x\, e  \, \f^{-1} \cD_a \f  \cD^a \f (\cD_0 \f + M^b  \cD_b \f ) \,.	
	\eea
}
\item {For the next three models, the scaling dimension of the scalar field $\f$ is $\o = \ft{z-d}{2}$.
\bea
S^{(9)}_{z\neq2} &=& \int dt\, d^d x\, e  \, \f \, \cD^a M_a  (\cD_0 \f + M^b \cD_b \f) \,,\nn\\
S^{(10)}_{z\neq2} &=& \int dt\, d^d x\, e  \, \f \triangle \f (M_0 + \ft12 M_a M^a) \,,\nn\\
S^{(11)}_{z\neq2} &=& \int dt\, d^d x\, e  \,  \cD_a \f  \cD^a \f (M_0 + \ft12 M_b M^b) \,.
\eea
}
\item {For the final kinetic term with a single derivative, the scaling dimension of the scalar field $\f$ is $\o = \ft{2z-d-2}{2}$
\bea
S^{(12)}_{z\neq2} &=& \int dt\, d^d x\, e  \, \f^2 \, \cD^a M_a  (M_0 + \ft12 M_b M^b) \,.
\eea}
	\end{enumerate}
}
\item {$n_t = 2:$ When we have two time derivative acting on $\f$ or $\chi$, the possible $z\neq2$ scale-invariant actions are
\bea
S^{(13)}_{z\neq2} &=& \int dt\, d^d x\, e \, \f \, (\cD_0^2 \f + 2 M^a \cD_0 \cD_a \f + M^a M^b \cD_a \cD_b \f ) \,,\nn\\
S^{(14)}_{z\neq2} &=& \int dt\, d^d x\, e \, \f^2 \, (\cD_0 M_0 + 2 M^a \cD_0 M_a + M^a M^b \cD_a M_b) \,,\nn\\
S^{(15)}_{z\neq2} &=& \int dt\, d^d x\, e \, \f^2 \, (M_0 + \ft12 M_a M^a)^2 \,.
\eea
Unlike the potentials and $n_t = 1$ models, we need to choose a different scaling dimension for each of the $n_t = 2$ actions due to the fact that the scaling dimension of $M_0$ is $z$-dependent, $\d M_0 = 2(1-z) \L_D M_0$. For $S^{(13)}_{z\neq2}$ we have $\o = \ft{z-d}{2}$, while for $S^{(14)}_{z\neq2}$ we have $\o = \ft{2z-d-2}{2}$ and for $S^{(15)}_{z\neq2}$ we have $\o = \ft{3z-d-4}{2}$.}
\end{itemize}
In our construction above, we avoid models that are equivalent to each other by means of partial integration or combination of other invariant actions, e.g. it is possible can also produce an invariant action using (\ref{DD0f}) and multiplying it with the Galilean-invariant covariant derivative $\cD^a \f$. However, such a model can be obtained by a partial integration of $S^{(9)}_{z\neq2}$. 

\subsubsection{Curvature Terms}
\paragraph{}
Following the $z=2$ discussion, we will now consider the curvature invariants. First, we enumerate the curvature invariants that are required for the commutation relations as mentioned above. The models that include the non-zero curvature $R_{\m\n}{}^a (G)$ are given by
\bea
S^{(16)}_{z\neq2} &=& \int dt\, d^d x\, e \, \f^2 \Big(\tau^\mu e^\nu{}_{a}R_{\mu\nu}{}^{a}(G) + M^b[2\tau^\mu e^\nu{}_{a} R_{\mu\nu}{}^{ab}(J) + M^c e^\mu{}_{c}e^{\nu}{}_{a} R_{\mu\nu}{}^{a}{}_b (J)]\Big)  \,,\nn\\
S^{(17)}_{z\neq2} &=& \int dt\, d^d x\, e \, \f \Big( R_{bc}(J) M^c + R_{ab}{}^a (G) \Big) \cD^b \f  \,.
\eea
For $S^{(16)}_{z\neq2}$, the scaling dimension of $\f$ is given by $\o = \ft{2-d}{2}$, while for $S^{(17)}_{z\neq2}$ we have $\o = \ft{z-d}{2}$. We also have the following two invariants that replaces the $z=2$ scale invariant action $S^{(21)}$ given in (\ref{RJSch}), in the case of $z\neq 2$ scale-extended non-relativistic gravity
\bea
S^{(18)}_{z\neq2} &=& \int dt\, d^d x\, e \, \f R(J)  (\cD_0 \f + M^a \cD_a \f)  \,,\nn\\
S^{(19)}_{z\neq2} &=& \int dt\, d^d x\, e \, \f^2 R(J)  (M_0 + \ft12 M_a M^a)  \,,
\eea
For $S^{(18)}_{z\neq2}$, the scaling dimension of $\f$ is given by $\o = \ft{2-d}{2}$, while for $S^{(19)}_{z\neq2}$ we have $\o = \ft{z-d}{2}$. Other curvature invariants that include the contraction of the rotation curvature $R_{abcd} (J)$ are the same as $z=2$ as given in (\ref{RJSch}) from $S^{(14)}$ to $S^{(20)}$, thus we will not give them here.

\subsection{$z\neq2$ Scale Invariance and the Ho\v{r}ava-Lifshitz Gravity}
\paragraph{}
In this section, our purpose is to combine the $z\neq2$ scale invariant gravity models that we constructed to identify the $z\neq2$ scale extension of the Ho\v{r}ava-Lifshitz gravity. Thus, we start this section with a brief review of the dictionary between the dynamical Newton-Cartan geometry and the Ho\v{r}ava-Lifshitz gravity that was put forward in \cite{Hartong:2015zia}. We refer to \cite{Hartong:2015zia} for readers interested in the details of the dictionary we review here.
\begin{enumerate}
\item {\textbf{Coordinates:} In order to define the Ho\v{r}ava-Lifshitz variables in terms of the fields in the scale-extended Newton-Cartan geometry, we first assume the hypersurface orthogonality condition
	\bea
	\t_{[\m} \partial_{\n} \t_{\r]} = 0 \,,
	\eea
which is satisfied by the $z\neq2$ scale invariant theory due to the constraint $R_{\m\n}(H) = 0$, see (\ref{RHSolve}). Next, we consider the $(d+1)$-dimensional ADM decomposition of the metric tensor
where metric tensor $g_{\m\n}$  that we defined in (\ref{gmn}). This leads to the following relations between the components of $h^{\m\n}, \hat{h}_{\m\n}, \t_{\m}, \hat{\t}^\m$ and the lapse function $N = N(t,{x})$, the shift vector $N^i = N^i(t,{x})$, and the $d$-metric $\g_{ab}$ \cite{Hartong:2015zia}
\bea
&& \t_t = N\,, \quad  \t_i = 0 \,,\nn\\
&& h^{tt} = h^{ti} = h^{it} = 0 \,, \quad h^{ij} = \g^{ij} \,,\nn\\
&&   \hat{\t}^t = N^{-1} \,, \quad \hat{\t}^i = - N^{-1} N^i \,,\nn\\
&& \hat{h}_{tt} = \g_{ij} N^i N^j \,,\quad \hat{h}_{ti} = \hat{h}_{it} = \g_{ij} N^j \,, \quad \hat{h}_{ij} = \g_{ij} \,, 
\label{ADMDecomp}
\eea
which implies the following expressions for $h_{\m\n}$ and $\t^\m$
\bea
&& \t^t = N^{-1} \,, \quad \t^i = 0	\,,\qquad h_{tt} = h_{ti} = h_{it} = 0 \,, \quad h_{ij} = \g_{ij} \,.
\eea
Here, it is important to note that we split the $\m$-index into coordinates $t$ and $x^i$. Using these relations, we also identify the $\rm U(1)$-invariant vector field $M_\m$ as
\bea
M_t = - \frac{1}{2N} \g_{ij} N^i N^j + \Phi N  \,, \qquad M_i = - \frac{1}{N} \g_{ij} N^j \,.
\eea
where $\Phi$ is the Newtonian potential that we defined in (\ref{Potential}). Finally, based on the twistless condition (\ref{Twistb}), we observe that this condition is fixed by $b_a = e^\m{}_a b_\m$ since the twistless condition imply
\bea
\partial_{[\m} \t_{\n]} = z b_{[\m} \t_{\n]} = z b_a e_{[\m}{}^a \t_{\n]}\,.
\eea 
Thus, in order to define the twistless condition in terms of the ADM variables, we define a vector, $a_\m$, as follows \cite{Hartong:2015zia}
\bea
a_\m = \cL_{\hat{\t}} \t_\m = \hat{\t}^\n (\partial_\n \t_{\m} - \partial_\m \t_{\n} ) = - z e_\m{}^a b_a \,,
\eea
where the last part of the equation is fixed by the twistless condition (\ref{Twistb}). Using the condition (\ref{Twistb}), we determine the vector $a_\m$ as  \cite{Hartong:2015zia}
\bea
a_t = N^i a_i \,, \qquad a_i = - N^{-1} \partial_i N \,.
\eea
Note that temporal component $b_0$ does not play a role in the connection between the Ho\v{r}ava-Lifshitz gravity and the Newton-Cartan geometry.}
\item {\textbf{Geometry:}
When we keep the lapse function only as a function of time only $N=N(t)$, we are dealing with the projectable Ho\v{r}ava-Lifshitz gravity. On the geometry side, $N=N(t)$ corresponds to the torsionless Newton-Cartan geometry since it gives rise to $\partial_{[\m} \t_{\n]} = 0$. When the lapse function $N$ is left arbitrary, we are dealing with the non-projectable Ho\v{r}ava-Lifshitz gravity, which corresponds to the twistless-torsional Newton Cartan geometry.
}
\item {\textbf{Curvatures:} The second fundamental form, or the extrinsic curvature is defined as
\bea
K_{ij} &=& \frac{1}{2N} \Big( \partial_t {\g}_{ij} - \bar\nabla_i N_j - \bar\nabla_j N_i \Big) \,,
\label{Kij}
\eea
where $\bar\nabla_i$ denotes the d-dimensional covariant derivative with respect to the d-metric $\g_{ij}$
\bea
\bar\nabla_i N_j = \partial_i N_j - \bar\G_{ij}^k N_k \,, 
\eea
where $\bar\G_{ij}^k$ denotes the components of the Christoffel connection for the d-metric $\g_{ij}$
\bea
\bar\G_{ij}^k  = \frac12 \g^{km} \Big( \partial_i \g_{jm} + \partial_j \g_{im} - \partial_m \g_{ij}  \Big) \,.
\eea
In order to relate the extrinsic curvature (\ref{Kij}) to the Newton-Cartan variables, we first make the following definition for the scale-covariant derivative of the $\U(1)$-invariant vector $M_a$ 
\bea
K_{ab}^\prime =  \cD_a M_b = \widetilde\nabla_{(a} M_{b)} + z b_{(a} M_{b)} - \d_{ab} b^c M_c - \d_{ab} b_0 \,.
\eea
where $\widetilde\nabla_a$ refers to the Galilean gauge-covariant piece of the scale covariant derivative
\bea
\widetilde\nabla_a M_b = e^\m{}_a \Big( \partial_\m M_b - \O_{\m b}{}^c M_c - \O_{\m b} \Big) \,.
\eea
Here, we also decomposed the rotation and boost gauge connections of the $z\neq2$ scale extended Bargmann algebra to the  of the Bargmann algebra as
\bea
\o_{\m}{}^{ab} &=& \O_{\m}{}^{ab} + 2 e_\m{}^{[a} b^{b]} - (z-2) e^{\nu a}e^{\rho b}\tau_\mu b_{[\n} m_{\r]} \,,\nn\\
\o_\m{}^a &=& \O_{\m}{}^{a}  +e_{\m}{}^a \tau^\nu b_\nu + (z-2) e^{\n a}  b_{[\m}m_{\n]} + (z-2) \tau_\m \tau^\rho e^{\nu a} b_{[\rho}m_{\n]}  
\eea
Here, we represent the rotation and boost gauge fields of the Bargmann algebra with $\O_{\m}{}^{ab}$ and $\O_{\m}{}^{a}$ in the respective order to distinguish these quantities with the relevant $z\neq2$ scale invariant ones. Based on our previous conclusion that $b_0$ does not play a role in the Ho\v{r}ava-Lifshitz gravity, it is best to consider models where $b_0$ drops out. This can be achieved in two ways
\begin{enumerate}[i.]
\item {We can decompose $K_{ab}^\prime$ as
\bea
K_{ab}^\prime = K_{ab} - \d_{ab} (b^c M_c +  b_0 ) \qquad \text{such that} \qquad K_{ab} = \widetilde\nabla_{(a} M_{b)} + z b_{(a} M_{b)} \,,
\eea
in which case the following combination has no $b_0$ term
\bea
K_{ab}^\prime K^{\prime ab} - \frac{1}{d} K^{\prime2} = K_{ab} K^{ab} - \frac{1}{d} K^{2} \,,
\eea
where $K \equiv K^a{}_a$ and $K^\prime \equiv K^{\prime a}{}_a$. This is precisely what was found as the kinetic term of conformal Ho\v{r}ava-Lifshitz gravity in \cite{Hartong:2015zia}.}
\item {We can consider the combination of $K_{ab}^\prime$ with the $z\neq2$ scale covariant combination (\ref{D0})
\bea
K^\prime_{ab} - \d_{ab} (\f^{-1} \cD_0 \f + M^a \f^{-1} \cD_a \f) \,,
\eea
in which case, as in the previous scenario, only the $K_{ab}$ part of $K_{ab}^\prime$ survives. }
\end{enumerate}
Therefore, we only need to worry about the relation between the $K_{ab}$ and $K_{ij}$ since the remaining terms can either be absorbed into the $z\neq2$ scale covariant combination (\ref{D0}), or can be canceled out by choosing a proper combination of $K_{ab}^\prime K^{ab \prime}$ and $K^{\prime2}$. As noted in \cite{Afshar:2015aku}, $K_{ab}$ can be written as
\bea
K_{ab} &=& e^\m{}_a e^\n{}_b \Big( \frac12 \cL_{\t} h_{\m\n} + \frac12 \nabla_\m (P_\n^\r M_\r )  + \frac12 \nabla_\n (P_\m^\r M_\r ) + z M_{(\m} b_{\n)} \Big)\,,
\eea
where 
\bea
\nabla_\m M_\n &=& \partial_\m M_\n - \G_{\m\n}^\r M_\r \,.
\eea
From this expression, we observe that one can write down the kinetic terms $K_{ab} K^{ab} = K_{ij} K^{ij}$ and $K_{a}{}^a = \g^{ij} K_{ij}$ upon using the map between the ADM variables and the Newton-Cartan fields (\ref{ADMDecomp}). }
\end{enumerate}
With these results in hand, we give the $z\neq2$ scale-extended Ho\v{r}ava-Lifshitz gravity as
\bea
S_{z\neq2}^{\rm HL} &=& S_{z\neq2}^{(3)} - \l \, S^{(2)}_{z\neq2} + S_{\mathcal V} \,,
\eea
where $\l$ is an arbitrary parameter and $S_\cV$ represents any remaining combination of actions that we constructed for $\chi$, $\f$ and group theoretical curvatures in the previous sections. 

%%%%%%%%%%%%%%%%%%%%%%%%%%%%%%%%%%%%%%%%%%%%%%%%%%%%%%%%%%%%%%%%%
\section{Conclusions}{\label{Conc}}
%%%%%%%%%%%%%%%%%%%%%%%%%%%%%%%%%%%%%%%%%%%%%%%%%%%%%%%%%%%%%%%%%
\paragraph{}
In this paper we present a detailed study on the construction of $z=2$ and $z\neq2$ scale invariant extension of the Ho\v{r}ava-Lifshitz gravity. To achieve these result, we developed a non-relativistic scale invariant tensor calculus and constructed scale invariant actions. Our results also enabled us to put an explicit distinction between the scale and Schr\"odinger invariance for $z=2$ non-relativistic gravity.

There are a number of directions to pursue following our work. First of all, the formulation we present here is not torsional since the gauging procedure we applied for the $z=2$ as well as the $z\neq2$ theories gave rise to a symmetric connection. In order to introduce a torsion, one can follow the idea presented in a recent work \cite{Bergshoeff:2017dqq} for the gauging of the Schr\"odinger algebra with torsion and repeat the construction procedure that we applied here. Furthermore, the scale or Schr\"odinger symmetry corresponds to a special choice of non-metricity in the compatibility equation, and it is possible to have a more general classification of non-relativistic geometries by imposing a more general non-metricity. This classification has been done for the relativistic scenarios in \cite{Jimenez:2015fva}, and it would be interesting to see the full classification of non-relativistic geometries with an arbitrary vector distortion. Another interesting direction concerns the supersymmetric completion of the Ho\v{r}ava-Lifshitz gravity. In \cite{Bergshoeff:2015ija} three-dimensional $\cN = 2$ Schr\"odinger supergravity was achieved by gauging the Schr\"odinger superalgebra. As the non-relativistic scalar multiplet of includes a complex scalar, it is possible to extend the Schr\"odinger tensor calculus of \cite{Afshar:2015aku} to a super-Schr\"odinger tensor calculus, which would than give rise to the three-dimensional $\cN=2$ Ho\v{r}ava-Lifshitz supergravity upon using the dictionary developed in \cite{Hartong:2015zia} and gauge fixing the redundant superconformal symmetries. Finally, the Schr\"odinger transformations are not true analogue of the relativistic conformal symmetry as they leave the action of a massive non-relativistic particle invariant. The true non-relativistic analogue of the relativistic conformal symmetry, which leaves the action of a massless non-relativistic particle is invariant is called the Galilean conformal algebra \cite{Bagchi:2009my}. It would interesting to see whether the non-relativistic tensor calculus can be extended to Galilean conformal algebra, and whether it is possible to distinguish a Galilean conformal gravity from a Galilean scale invariant gravity.

%%%%%%%%%%%%%%%%%%%%%%%%%%%%%%%%%%%%
\section*{Acknowledgment}
%%%%%%%%%%%%%%%%%%%%%%%%%%%%%%%%%%%%
\paragraph{}
We thank Hamid Afshar, Eric Bergshoeff, Jelle Hartong, Kristan Jensen and Jan Rosseel for useful discussions and clarifications on non-relativistic gravity. The authors would like to thank the referees for insightful comments which helped to improve this paper. DOD is supported by Basic Science Research Program through the National Research Foundation of Korea (NRF) funded by the Ministry of Education, Science and Technology (2016R1A2B401304). NO and UZ are supported in parts by Istanbul Technical University Research Fund under grant number TDK-2018-41133.
%%%%%%%%%%%%%%%%%%%%%%%%%%%%%%%%%%%% 

\providecommand{\href}[2]{#2}\begingroup\raggedright\endgroup


\begin{thebibliography}{99}

\bibitem{Cartan1}
E. Cartan,
``Sur les variétés à connexion affine et la théorie de la relativité généralisée. (première partie),''
Annales Sci.Ecole Norm.Sup. 40 (1923) 325-412

\bibitem{Cartan2}
E. Cartan,
``Sur les variétés à connexion affine et la théorie de la relativité généralisée. (première partie) (Suite),''
Annales Sci.Ecole Norm.Sup. 41 (1924) 1-25

	\bibitem{Son:2013rqa}
D.~T. Son, ``{Newton-Cartan Geometry and the Quantum Hall Effect},''
\href{http://www.arXiv.org/abs/1306.0638}{{\tt 1306.0638}}.
%%CITATION = ARXIV:1306.0638;%%.

\bibitem{Gromov:2014vla}
A.~Gromov and A.~G. Abanov, ``{Thermal Hall Effect and Geometry with
	Torsion},'' {\em Phys. Rev. Lett.} {\bf 114} (2015) 016802,
\href{http://www.arXiv.org/abs/1407.2908}{{\tt 1407.2908}}.
%%CITATION = ARXIV:1407.2908;%%.

\bibitem{Geracie:2014nka}
M.~Geracie, D.~T. Son, C.~Wu, and S.-F. Wu, ``{Spacetime Symmetries of the
	Quantum Hall Effect},'' {\em Phys. Rev.} {\bf D91} (2015) 045030,
\href{http://www.arXiv.org/abs/1407.1252}{{\tt 1407.1252}}.
%%CITATION = ARXIV:1407.1252;%%.

\bibitem{Moroz:2014ska}
S.~Moroz and C.~Hoyos, ``{Effective theory of two-dimensional chiral
	superfluids: gauge duality and Newton-Cartan formulation},'' {\em Phys. Rev.}
{\bf B91} (2015), no.~6, 064508,
\href{http://www.arXiv.org/abs/1408.5911}{{\tt 1408.5911}}.
%%CITATION = ARXIV:1408.5911;%%.

\bibitem{Horava1}
 P.~Horava,
``Quantum Gravity at a Lifshitz Point,''
Phys.\ Rev.\ D {\bf 79}, 084008 (2009)
\href{https://arxiv.org/abs/0901.3775}{{\tt 0901.3775}}.

\bibitem{Horava2}
P.~Horava,
``Membranes at Quantum Criticality,''
JHEP {\bf 0903}, 020 (2009)
\href{https://arxiv.org/abs/0812.4287}{{\tt 0812.4287}}.

\bibitem{Leiva:2003kd}
C.~Leiva and M.~S. Plyushchay, ``{Conformal symmetry of relativistic and
	nonrelativistic systems and Ads / CFT correspondence},'' {\em Annals Phys.}
{\bf 307} (2003) 372--391,
\href{http://www.arXiv.org/abs/hep-th/0301244}{{\tt hep-th/0301244}}.
%%CITATION = HEP-TH/0301244;%%.

\bibitem{Balasubramanian:2008dm}
K.~Balasubramanian and J.~McGreevy, ``{Gravity duals for non-relativistic
	CFTs},'' {\em Phys. Rev. Lett.} {\bf 101} (2008) 061601,
\href{http://www.arXiv.org/abs/0804.4053}{{\tt 0804.4053}}.
%%CITATION = ARXIV:0804.4053;%%.

\bibitem{Son:2008ye}
D.~T. Son, ``{Toward an AdS/cold atoms correspondence: A Geometric realization
	of the Schrodinger symmetry},'' {\em Phys. Rev.} {\bf D78} (2008) 046003,
\href{http://www.arXiv.org/abs/0804.3972}{{\tt 0804.3972}}.
%%CITATION = ARXIV:0804.3972;%%.

\bibitem{Herzog:2008wg}
C.~P. Herzog, M.~Rangamani, and S.~F. Ross, ``{Heating up Galilean
	holography},'' {\em JHEP} {\bf 11} (2008) 080,
\href{http://www.arXiv.org/abs/0807.1099}{{\tt 0807.1099}}.
%%CITATION = ARXIV:0807.1099;%%.

\bibitem{Duval:2008jg}
C.~Duval, M.~Hassaine, and P.~A. Horvathy, ``{The Geometry of Schrodinger
	symmetry in gravity background/non-relativistic CFT},'' {\em Annals Phys.}
{\bf 324} (2009) 1158--1167,
\href{http://www.arXiv.org/abs/0809.3128}{{\tt 0809.3128}}.
%%CITATION = ARXIV:0809.3128;%%.

\bibitem{Kachru:2008yh}
S.~Kachru, X.~Liu, and M.~Mulligan, ``{Gravity duals of Lifshitz-like fixed
	points},'' {\em Phys. Rev.} {\bf D78} (2008) 106005,
\href{http://www.arXiv.org/abs/0808.1725}{{\tt 0808.1725}}.
%%CITATION = ARXIV:0808.1725;%%.

\bibitem{Taylor:2008tg}
M.~Taylor, ``{Non-relativistic holography},''
\href{http://www.arXiv.org/abs/0812.0530}{{\tt 0812.0530}}.
%%CITATION = ARXIV:0812.0530;%%.

\bibitem{Wang:2017brl} 
A.~Wang,
``Horava gravity at a Lifshitz point: A progress report,''
Int.\ J.\ Mod.\ Phys.\ D {\bf 26}, no. 07, 1730014 (2017)
\href{http://www.arXiv.org/abs/1701.06087}{{\tt 1701.06087}}.

\bibitem{Hartong:2015zia}
J.~Hartong and N.~A.~Obers,
``Horava-Lifshitz gravity from dynamical Newton-Cartan geometry,''
JHEP {\bf 1507}, 155 (2015)
\href{https://arxiv.org/abs/1504.07461}{{\tt 1504.07461}}.

\bibitem{Afshar:2015aku} 
H.~R.~Afshar, E.~A.~Bergshoeff, A.~Mehra, P.~Parekh and B.~Rollier,
``A Schr\"odinger approach to Newton-Cartan and Hořava-Lifshitz gravities,''
JHEP {\bf 1604}, 145 (2016)
\href{https://arxiv.org/abs/1512.06277}{{\tt 1512.06277}}.

\bibitem{Bergshoeff:2014uea}
E.~A.~Bergshoeff, J.~Hartong and J.~Rosseel,
``Torsional Newton - Cartan geometry and the Schr\"odinger algebra,''
Class.\ Quant.\ Grav.\  {\bf 32}, no. 13, 135017 (2015)
\href{https://arxiv.org/abs/1409.5555}{{\tt 1409.5555}}.

\bibitem{Jensen:2014aia}
K.~Jensen,
``On the coupling of Galilean-invariant field theories to curved spacetime,''
\href{https://arxiv.org/abs/1408.6855}{{\tt 1408.6855}}.


\bibitem{Andringa:2010it}
R.~Andringa, E.~Bergshoeff, S.~Panda and M.~de Roo,
``Newtonian Gravity and the Bargmann Algebra,''
Class.\ Quant.\ Grav.\  {\bf 28}, 105011 (2011)
\href{https://arxiv.org/abs/1011.1145}{{\tt 1011.1145}}.

\bibitem{Dautcourt}
G.~Dautcourt,
``On the Newtonian limit of General Relativity,''
Acta.~Phys.~Pol.~B {\bf 21} (1990) 766-765.


\bibitem{Trautman}
A.~Trautman,
``Sur la theorie newtonienne de la gravitation,'' Compt. ~Rend. ~Acad. ~Sci. ~Paris 247 (1963) 617. 

\bibitem{Ehlers}
J.~Ehlers,
``\"{U}ber den Newtonschen Grenzwert,''
in {\sl Grundlagen-probleme der modernen Physik},
ed. J. Nitsch, J. Pfarr and E.-W. Stachow, 
Bibliographisches Institut Mannheim/Wien/Z\"urich
(1981).


\bibitem{Christensen:2013lma}
M.~H.~Christensen, J.~Hartong, N.~A.~Obers and B.~Rollier,
``Torsional Newton-Cartan Geometry and Lifshitz Holography,''
Phys.\ Rev.\ D {\bf 89}, 061901 (2014)
\href{https://arxiv.org/abs/1311.4794}{{\tt 1311.4794}}.

\bibitem{Banerjee:2016laq} 
R.~Banerjee and P.~Mukherjee,
``Torsional Newton–Cartan geometry from Galilean gauge theory,''
Class.\ Quant.\ Grav.\  {\bf 33}, no. 22, 225013 (2016)
\href{https://arxiv.org/abs/1604.06893}{{\tt 1604.06893}}.

\bibitem{Hartong:2015wxa}
J.~Hartong, E.~Kiritsis and N.~A.~Obers,
``Field Theory on Newton-Cartan Backgrounds and Symmetries of the Lifshitz Vacuum,''
JHEP {\bf 1508}, 006 (2015)
\href{https://arxiv.org/abs/1502.00228}{{\tt 1502.00228}}.

\bibitem{Mitra:2015twa} 
A.~Mitra,
``Nonrelativistic fluids on scale covariant Newton–Cartan backgrounds,''
Int.\ J.\ Mod.\ Phys.\ A {\bf 32}, no. 36, 1750206 (2017)
\href{https://arxiv.org/abs/1508.03207}{{\tt 1508.03207}}.

\bibitem{Bergshoeff:2017dqq} 
E.~Bergshoeff, A.~Chatzistavrakidis, L.~Romano and J.~Rosseel,
``Newton-Cartan Gravity and Torsion,''
JHEP {\bf 1710}, 194 (2017)
\href{https://arxiv.org/abs/1708.05414}{{\tt 1708.05414}}.

\bibitem{Jimenez:2015fva} 
J.~Beltran Jimenez and T.~S.~Koivisto,
``Spacetimes with vector distortion: Inflation from generalised Weyl geometry,''
Phys.\ Lett.\ B {\bf 756}, 400 (2016)
\href{https://arxiv.org/abs/1509.02476}{{\tt 1509.02476}}.

\bibitem{Bergshoeff:2015ija} 
E.~Bergshoeff, J.~Rosseel and T.~Zojer,
``Newton-Cartan supergravity with torsion and Schr\"odinger supergravity,''
JHEP {\bf 1511}, 180 (2015)
\href{https://arxiv.org/abs/1509.04527}{{\tt 1509.04527}}.

\bibitem{Bagchi:2009my}
A.~Bagchi and R.~Gopakumar, ``{Galilean Conformal Algebras and AdS/CFT},'' {\em
	JHEP} {\bf 07} (2009) 037,
\href{http://www.arXiv.org/abs/0902.1385}{{\tt 0902.1385}}.

\end{thebibliography}
\end{document}